\documentstyle[12pt]{article}
\renewcommand{\theequation}{\arabic{section}.\arabic{equation}}

\begin{document}
\begin{titlepage}
\begin{center}
\vspace{1cm}
\hfill
\vbox{
    \halign{#\hfil         \cr
           hep-th/9803162 \cr
           IPM-98-284\cr
           March 1998\cr} }
      
\vskip 1cm
{\Large \bf
Distributed Systems of Intersecting\\ 
Branes at Arbitrary Angles}
\vskip 0.5cm
{\bf R. Abbaspur$^{*, \dagger}$\footnote{e-mail:abbaspur@netware2.ipm.ac.ir} 
and
H. Arfaei$^{*, \dagger}$\footnote{e-mail:arfaei@theory.ipm.ac.ir}}\\ 
\vskip .25in
{\em
$^*$Institute for Studies in Theoretical Physics and Mathematics,  \\
P.O. Box 19395-5531,  Tehran,  Iran.\\
$^\dagger$Department of Physics,  Sharif University of Technology, \\
P. O. Box 19365-9161,  Tehran,  Iran.}
\end{center}
\vskip 0.5cm

\begin{abstract}
A `reduced' action formulation for a general 
class of the supergravity solutions, 
corresponding to the  
`marginally' bound `distributed' systems of 
various types of branes at arbitrary angles, is developed. 
It turns out that all the information regarding 
the classical features of such solutions is encoded 
in a first order Lagrangian 
(the `reduced' Lagrangian) corresponding to the 
desired geometry of branes. The marginal solution for a system of $N$ 
such distributions (for various distribution functions) span an $N$   
dimensional submanifold of the fields' configuration (target) space,
 parametrised by a set of $N$ independent harmonic functions
on the transverse space. This submanifold, which we call it as the 
`$H$-surface', is a null surface with 
respect to a metric on the configuration space, which is defined  
by the reduced Lagrangian. The equations of motion then 
transform to a set of equations describing the
embedding of a null geodesic surface 
in this space, which is identified as the $H$-surface. 
Using these facts, we present a very simple derivation of the
conventional orthogonal solutions together with their
intersection rules. Then a new solution 
for a (distributed) pair of $p$-branes at 
SU(2) angles in $D$ dimensions is derived.

\end{abstract}

\end{titlepage}\newpage

\def\pb{$p$-brane }
\def\pbs{$p$-branes }
\def\ff{form-field }
\def\ffs{form-fields }
\def\fp{form-potential }
\def\fps{form-potentials }
\def\ir{intersection rule }
\def\irs{intersection rules }
\def\eom{equation of motion }
\def\eoms{equations of motion }
\def\nfc{no-force conditions }

\def\A{\parallel}
\def\B{\perp}
\def\w{\wedge}
\def\W{\bigwedge}
\def\2{{1\over 2}}
\def\ra{\rightarrow}
\def\={\equiv}
\def\p{\partial}
\def\pl{\partial_\lambda}
\def\plp{\partial_{\lambda '}}
\def\pls{\partial_{\lambda ''}}
\def\pld{\dot{\partial}_\lambda}
\def\plpd{\dot{\partial}_{\lambda '}}

\def\a{\alpha}
\def\b{\beta}
\def\c{\gamma}
\def\d{\delta}
\def\e{\epsilon}
\def\ve{\varepsilon}
\def\g{\gamma}
\def\v{\upsilon}
\def\th{\theta}
\def\vth{\vartheta}
\def\k{\kappa}
\def\kp{\kappa '}
\def\ks{\kappa ''}
\def\l{\lambda}
\def\lp{\lambda '}
\def\ls{\lambda ''}
\def\m{\mu}
\def\n{\nu}
\def\o{\omega}
\def\x{\xi}
\def\r{\rho}
\def\vr{\varrho}
\def\s{\sigma}
\def\t{\tau}
\def\f{\phi}
\def\vf{\varphi}
\def\G{\Gamma}
\def\D{\Delta}
\def\L{\Lambda}
\def\O{\Omega}

\def\dxi{(dx^i)}
\def\dx{dx^{\mu}}
\def\dxp{dx_{\mu}}
\def\dy{dy_1^{m}}
\def\dyp{dy_2^{m'}}
\def\dz{dz^a}
\def\Xaz{X_1(z)}
\def\Xbz{X_2(z)}

\def\cA{{\cal  A}}
\def\cB{{\cal  B}}
\def\cF{{\cal  F}}
\def\cJ{{\cal  J}}
\def\cL{{\cal  L}}
\def\cM{{\cal  M}}
\def\cQ{{\cal  Q}}
\def\LF{{\cal L}_F}
\def\LG{{\cal L}_G}
\def\cH{{\cal H}}

\def\fa{\phi^\alpha}
\def\fb{\phi^\beta}
\def\fc{\phi^\gamma}
\def\rl{\rho_\lambda}
\def\rlp{\rho_\lambda '}
\def\rlz{\rho_\lambda (z)}
\def\rlzp{\rho_\lambda (z')}
\def\Xl{X_\lambda}
\def\Xlp{X_{\lambda '}}
\def\Xlz{X_\lambda (z)}
\def\Xlpz{X_{\lambda '}(z)}
\def\Hl{H_\lambda}
\def\Hlp{H_{\lambda '}}
\def\Hlz{H_\lambda (z)}
\def\Hlpz{H_{\lambda '}(z)}
\def\Fl{F_\lambda}
\def\Flp{F_{\lambda '}}
\def\hl{h_\lambda}
\def\hlp{h_{\lambda '}}
\def\hlz{h_\lambda (z)}
\def\hlpz{h_{\lambda '}(z)}
\def\fla{f_{\lambda\alpha}}
\def\flpa{f_{\lambda '\alpha}}
\def\ga{g_\alpha}
\def\xla{\xi_\lambda^\alpha}
\def\xlpa{\xi_{\lambda '}^\alpha}
\def\xaa{\xi_1^\alpha}
\def\xba{\xi_2^\alpha}
\def\Oab{\Omega_{\alpha\beta}}
\def\Obc{\Omega_{\beta\gamma}}
\def\Ohab{\hat{\Omega}_{\alpha\beta}}
\def\sa{{\sigma}_{\alpha}}
\def\bfl{{\bf f}_\lambda}
\def\bflp{{\bf f}_{\lambda '}}
\def\bg{{\bf g}}
\def\bO{{\bf\Omega}}
\def\xl{{\xi\kern -.43em\xi}_\lambda}
\def\xlp{{\xi\kern -.43em\xi}_{\lambda '}}
\def\Db{{\bar D}}
\def\dt{{\tilde d}}
\def\dtt{({\tilde d}+1)} 
\def\al{\alpha_\lambda}
\def\alp{\alpha_{\lambda '}}
\def\kl{\kappa_\lambda}
\def\klp{\kappa_{\lambda '}}
\def\Ml{M_\lambda}
\def\Ql{Q_\lambda}
\def\dl{d_\lambda}
\def\dlp{d_{\lambda '}}
\def\dtl{{\tilde d}_\lambda}
\def\dtlp{{\tilde d}_{\lambda '}}
\def\dta{{\tilde d}_1}
\def\dtb{{\tilde d}_2}
\def\aa{\alpha_1}
\def\ab{\alpha_2}
\def\ka{\kappa_1}
\def\kb{\kappa_2}
\def\iz{\int d^{{\tilde d}+2}z}
\def\izp{\int d^{{\tilde d}+2}z'}
\def\Baz{B_0(z)}
\def\Bbz{B_1(z)}
\def\Bcz{B_2(z)}
\def\dllp{\delta_{\lambda\lambda '}}
\def\Ai{A_{(i)}}
\def\Aj{A_{(j)}}
\def\Ari{A_{(i)}^r}
\def\Arj{A_{(j)}^r}
\def\hij{h_{ij}}
\def\hijp{h^{ij}}
\def\gij{\gamma_{ij}}
\def\gmm'{\gamma_{mm'}}
\def\gijp{\gamma^{ij}}
\def\gmmp{\gamma_{mm'}}
\def\gmnp{\gamma_{mn'}}
\def\grs{\gamma_{rs}}
\def\eij{\eta_{ij}}
\def\emn{\eta_{\m\n}}
\def\dmn{\delta_{mn}}
\def\dmpnp{\delta_{m'n'}}
\def\ul{u_{\lambda\lambda '}}
\def\Ul{U_{\lambda\lambda '}}
\def\ulr{u_{\lambda\lambda '}^r}
\def\clri{c_\lambda^{r(i)}}
\def\clrj{c_\lambda^{r(j)}}
\def\el{\epsilon_\lambda}
\def\eli{\epsilon_{\lambda (i)}}
\def\elj{\epsilon_{\lambda (j)}}
\def\fl{f_{\lambda\lambda '}}
\def\Ll{\cal l{L}_{\lambda\lambda '}}
\def\cli{c_\lambda^{(i)}}
\def\clj{c_\lambda^{(j)}}
\def\clpi{c_{\lambda '}^{(i)}}
\def\clpj{c_{\lambda '}^{(j)}}
\def\cllp{c_{\lambda\lambda '}^{r(i)}}
\def\clpl{c_{\lambda '\lambda}^{r(i)}}
\def\ori{\omega_{(i)}^r}
\def\cllpri{c_{\lambda\lambda '}^{r(i)}}
\def\Flri{F_\lambda^{r(i)}}
\def\flrs{f_{\l (rs)}}
\def\Fli{F_\lambda^{(i)}}
\def\clm{c_\lambda^{(\m)rs}}
\def\clrs{c_\lambda^{rs}}
\def\chlrs{{\hat c}_\lambda^{rs}}
\def\ch{{\hat c}}
\def\Flrs{F_\lambda^{rs}}
\def\bH{H}
\def\oa{\omega_\alpha}
\def\ob{\omega_\beta}
\def\bet{\eta\kern -.48em\eta}
\def\bga{\gamma\kern -.5em\gamma}
\def\bo{\omega\kern -.58em\omega}
\def\Ulf{U_{\lambda\lambda '}({\bf\phi}, H)}
\def\ulf{u_{\lambda\lambda '}({\bf\phi})}
\def\Oabf{\Omega_{\alpha\beta}({\bf\phi})}
\def\HH{\partial H_\lambda .\partial H_{\lambda '}}
\def\gg{2(G+\alpha_r\phi )}
\def\bh{{\bf h}}
\def\h{(h_{ij})}
\def\hp{(h^{ij})}
\def\tr{Tr({\bf h}^{-1}\partial{\bf h})}
\def\trb{Tr^2({\bf h}^{-1}\partial{\bf h})}
\def\bff{{\bf \phi}}
\def\Ar{{\cal A}^r}
\def\Fr{{\cal F}^r}
\def\st{sin\theta}
\def\sbt{sin^2\theta}
\def\ct{cos\theta}
\def\cbt{cos^2\theta}
\def\ea{\eta^\alpha}
\def\bX{{\bf X}}
\def\pad{{\dot{\partial}}_1}
\def\pbd{{\dot{\partial}}_2}
\def\bOh{\hat{\bf\Omega}}
\def\bs{\sigma\kern -.53em\sigma}

\def\ft{\footnote}
\def\nn{\nonumber}
\def\be{\begin{equation}}
\def\ee{\end{equation}}
\def\bea{\begin{eqnarray}}
\def\eea{\end{eqnarray}}
\def\np{\newpage}

{\Large{\bf Introduction}}\\
BPS configurations of intersecting branes have been the basis of many recent
developements in string and M-theory. 
This is mainly because of their essential role in establishing
string dualities, counting the entropy of extremal black 
holes and constructing supersymmetric gauge theories (for reviews see 
\cite{14}-\cite{40} and referrences therein). 
The construction of supergravity
solutions corresponding to such configurations of branes have been
implemented using four main approaches. 
These include: supersymmetry 
techniques \cite{4,1,3,25,7,39},  
duality transformations \cite{32,38,23,7,39}, 
dimensional reduction 
and oxidation \cite{20,12,7,39} and lastly direct solving of
the bosonic field equations of supergravity\cite{13,5,6,11,15,16,27}. 
If we are going to study, in a unified manner, 
a wide class of these solutions in various
types of $D$-dimensional supergravities, 
containing all types of \pbs ,  the only 
systematic approach is to use the direct method. 
In this approach one 
considers a purely bosonic theory consisting
of two sectors, the  `gravitational 
sector'  including a metric tensor and several
dilatons,  and the  `form-fields sector'  
consisting of antisymmetric tensors ( `form-fields' ) 
of various degrees. Such a model can be imagined to be the bosonic sector of
a supergravity theory in $D$ dimensions,  obtained from the 11D-theory via the
processes of dimensional reduction 
and truncation \cite{20}. In the usual approach, 
one writes the {\it complete set} of the field equations for the corresponding
$D$-dimensional action,  and tries to solve them by inserting a set of ansatze 
appropriate to the given configuration 
of \pbs . In spite of its logical completeness
and reliability,  such an approach involves extra complications due to the
appearance of coupled  Einstein-`form-fields' equations. 
Typically,  introducing an ansatz with a set of field equations, 
means  `reducing'  the  `actual'  
degrees of freedom (including spacetime dependences)
to the  `relevant'  (i.e. excited) ones 
describing the physical situation of interest. So an alternative
approach is to form a  `reduced action'  
describing only dynamics of the excited
degrees of freedom,  and try to 
solve their equations of motion. However, in this
manner we will lose the equations of 
motion corresponding to `irrelevant'  degrees
of freedom,  which appear as 
the constraints between relevant ones. For this reason,  not every
solution of the  `reduced theory'  is also a solution of the original one, 
but it is easy to see that the converse is true. 
The situation is different when certain
 `subspaces'  of the  
 `configuration space'  are concerned\ft{This is comparable
to a similar fact in the differential geometry that:
Every geodesic line of a 
Riemannian space,  lieing on an arbitrary surface,  is
also a geodesic line of the 
surface itself,  while the converse is not necessarily true. 
However for certain subspaces called  `geodesic surfaces'  the inverse theorem
is also true.}.
It turns out that supergravity BPS saturated solutions describe two types of
brane systems \cite{38,18}: $i$)marginal 
configurations and $ii$)non-marginal configurations.
In the first class the binding 
energy of the bound system of branes for their
{\it arbitrary separations} vanishes so that they don't feel any total force.
In the second class the binding energy is a negative quantity and the total
force between each pairs of branes is attractive. As a result the bound state
can be stable only at zero separations and the corresponding solutions are
spherically symmetric around 
the common centre of branes in their transverse space.
The marginality condition in the 
first class,  imposes heavy constraints on the form of the
corresponding solutions. They 
must satisfy certain  `extremality'  and  `no-force' 
conditions. Further,  all the 
field variables must be written totally as functions 
of a set of  `independent'  harmonic functions \cite{26}. It will be seen in
this paper that,  the (first order)  
`reduced Lagrangian'  of the theory $\cL$,  on the
 `subspace'  $H$ of such solutions,  
 identically vanishes. This property, with some
insights from the differential geometry,  enforces the idea that every
solution of the reduced 
theory on the  `$H$-surface'  must describe a solution of
the original theory as well. 
It seems that all the physical information regarding these
solutions comes essentially through the corresponding reduced actions.
In this paper we try to elaborate this idea,  to the extent
that it can be used for determining the 
solutions corresponding to the orthogonal, 
as well as the 
{\it non-orthogonal configurations} of \pbs . We will study solutions
with uniform distributions 
of branes along their relative transverse directions, 
but with arbitrary distributions 
along their overall transverse directions. We shall
refer to such systems as the 
`distributed' systems, which are described in terms of a set of
density functions in this paper. Continuous distributions in a low-energy
model,  can be interpreted as 
the long-distance limits of periodic (or un-periodic)
arrays of branes with relative 
separations of order $\sqrt{\a '}$ in the high-energy
string theory. In the low-energy limit the Kaluza-Klein modes corresponding
to the compactified relative 
transfers coordinates,  are averaged and the solutions
will depend only on the overall transverse coordinates \cite{35}.
We begin our study of these solutions in section 1, 
by a reformulation of the problem of two
different orthogonal branes,  
in the framework of a reduced theory. We will introduce
a set of constraints and insist 
on their determining role in solving the field
equations of the reduced theory. 
We will find in this approach,  in agreement 
with those of \cite{11,15},  
that the field equations are completely replaced with
a set of {\it algebraic constraints} 
determining parameters of the solution. The  
relations between couplings and dimensions 
emerge as the consistency conditions 
of these algebraic constraints. 
As a byproduct,  a Diophantine equation governing
the allowed marginal 
intersections of super \pbs is obtained. It is argued that
the usual relation between 
the dilaton couplings and the corresponding form-fields
degrees in supergravity \cite{13},  
is a direct result of supersymmetry. In section 2
we generalize the ideas of section 1 
to a system of multiply intersecting orthogonal 
branes. It will be seen that the 
algebraic constraints in matrix representation
have a universal form. Then the solution for these constraints together with 
their consistency conditions in compact forms will be given. In section 3 we
lay the foundations of a more general theory of the marginal brane
solutions including systems of 
branes at angles. General ansatze for the form-fields
describing purely electric-type branes,  
in terms of the ansatz for the metric
are derived. It is shown that these 
ansatze are dependent on a set of  `structure
constants' ,  which encode 
the geometry (i.e. angles) of the brane system. A general 
formula for these constants,  
in terms of the asymptotic form of the metric is
derived. The concepts of the  
`$H$-surface'  and an  `$H$-basis'  are defined. A
general formula for the reduced Lagrangian (RL),  describing such generalized
systems is derived. 
Using this RL the equations for the embedding of the  `$H$-surface'  
in the configuration space of the field variables will be written. Heuristic
derivations for the generalized 
versions of the constraints in sections 1 $\&$ 2
are given and the role of the $\cL =0$ equation is emphasized. In section 4
we apply the general 
framework of section 3,  to the problem of arbitrary branes at
angles. Then explicit solutions 
for the marginal configuration: $p\cap p=(p-2)$ using
this formulation is derived. During the way of this derivation we encounter
an integrability condition,  
from which the relation between angles (besides
other information) is obtained. 
In section 5 a formulation of the no-force
conditions appropriate to the framework of 
section 3 will be given. It is
shown that these conditions in combination with the extremality conditions
gives rise to a class of constraints,  
derived earlier on an ad hoc basis. The various (massless)
fields contributions to the static long-range 
potential (between a brane system
and a brane probe) are separated. 
The consistency conditions of section 1 emerge
as the balance conditions 
among the long-range forces. 
In section 6 the formulas
for masses and charges 
are presented and shown,  as is expected that,  these become 
proportional to the 
integrals of the density functions. In Appendix A a general
first order Lagrangian for gravity is formulated and its application for the
derivation of the RL's is indicated. In Appendix B we present two methods for
classifying the solutions of the Diophantine equation for intersections for
arbitrary spacetime dimension $D$,  and then give the classifications for
$D=4, 6, 10, 11$. In Appendix C the angles between a pair of branes  
in terms of the asymptotic form of the spacetime metric are
 defined. Finally in 
Appendix D we give the proofs of some $H$-surface identities.
In the following sections,  
we will use a model theory with an action of the form:
\be
I_D=\int d^Dx\left 
(\sqrt{-g}R-\2(\nabla\vf)^2-\2\sum_{r=1}^ne^{2{\a}_r\vf}{\cF}_r^2\right )
\label{I.1}\ee
where ${\cF}_r$'s 
denote $(p_r+2)$-form field-strengths defined as: ${\cF}_r=d{\cA}_r$, 
and ${\cF}_r^2\;\=\;{1\over{(p_r+2)!}}
F_{rM_1...M_{p_r+2}}F_r^{M_1...M_{p_r+2}}$.
Upon fixing the non-vanishing form-fields and the respective couplings 
${\a}_r$, we are able to handle various supergravity models with suitable 
truncations.\np

\section{Two Orthogonal Branes}
\setcounter{equation}{0}
To show the essence of the reduced theory 
formulation, we begin our discussion 
by concentrating  on a simple example.
We consider a simple distributed system constructed on the basis of a
pair of orthogonal \ft{Clearly the parallel situation $d_1\subseteq d_2$ is 
a special case where $\d =d_1$.} $(d_1-1, d_2-1)$-branes 
intersecting (overlapping)
\ft{In this paper we use the 
term  `intersecting branes'  in equal footing with 
 `overlapping branes', 
 i.e. different branes may have separated centres in their 
transverse space. 
We use the $world-volume$ notation: $d_1\cap d_2=\d$ for indicating such
intersections.} over
$(\d -1)$ dimensions in the 
spacetime of arbitrary dimension $D$. The (minimum)
dimension of a $homogeneous$ hyperplane filled with the 
$uniform$ (parallel) distributions of the two branes obviously is
\be        
d=d_1+d_2-{\d}                                                                                                           
\label{1.1} \ee
The overall
transverse space has 
$D-d\= {\dt} +2$ dimensions. We will use a set of Cartesian
coordinates $(x^{\m}, y_1^m, y_2^{m' }, z^a)$ as defined in table (1).
$$
\begin{array}{|c|c|l|}\hline
coordinate&dimension&tangent\;\; (v)\\ \hline
x^\m&\d&v\A d_1\,  , \, v\A d_2 \\
y_1^m&{\d}_1\= d_1-\d&v\A d_1\,  , \, v\B\d \\
y_2^{m' }&{\d}_2\= d_2-\d&v\A d_2\,  , \, v\B\d \\
z^a&\dt +2&v\B d_1 , v\B d_2\\  \hline
\end{array} \;\;\;\;\;\;\; table(1)
$$               
By construction all the components 
of any physical field for such a configuration
will depend only on the coordinates $(z^a)$ of the transverse space. We can
also allow a  `transverse'  
distribution of branes,  which unlike
the  `longitudinal'  
distributions may be inhomogeneous. Throughout this paper we
will describe such
 `transverse'  distributions 
 using the density functions ${\r}_{\l}(z)$,  where ${\l}$
labels\ft{In this paper 
we will use the indices $\l , \lp$, etc. and $\k , \kp$, etc.  
exclusively for the distribution quantities.}
various distributions in the problem (e.g. ${\l}=1, 2$ for now).
It must be remarked here that 
such distributions are possible only when each pair of
the parallel constituent 
branes can form stable bound states with zero binding
energy. It is widely
believed that this  `marginal'  binding property,  is a
characteristic of 
supersymmetric bound states \cite{4,3,14,19}. However it will become
clear during this section that 
the distributions of non-supersymmetric $p$-branes
in the non-supersymmetric 
theories are also possible. Also,  we will see that bound
states of various such distributions become possible whenever they
have a suitable number of common directions.\\
To construct field theoretic 
classical solutions,  as usual a set of ansatze for
the metric and the form 
fields are required. The ansatz for the metric is fixed
by means of its isometries 
defined by the set of commuting 
Killing vectors: $({\p}_{\m} , {\p}_m , {\p}_{m' })$, 
and its isotropies: $SO({\d} -1, 1)\times SO({\d}_1)\times SO({\d}_2)\times SO({\dt} +2)$.
Here we have demanded Lorentz invariance along the common 
$(x^{\m})$ 
directions. This requirement is believed to be a characteristic of super
symmetric solutions \cite{13,14}. 
However the solutions discussed here are not limited
to supersymmetric theories. 
Hence we write the ansatz for the metric as
\be   
ds^2=e^{2B_0(z)}dx^{\m} dx_{\m} 
+e^{2B_1(z)}dy_1^mdy_1^m+e^{2B_2(z)}dy_2^{m' }dy_2^{m' }
         + e^{2C(z)}dz^adz^a
\label{1.2} \ee
where $dx^{\m} dx_{\m} =\eta_{{\m}{\n}}
dx^{\m} dx^{\n}$ and $\eta_{{\m}{\n}}=diag(-1, +1, ..., +1)$.
We should fix the ansatze 
for the form fields as well. For this purpose we must specify
that whether the $(d_1-1, d_2-1)$-branes carry electric or magnetic charges, 
i.e. whether they couple to $(d_1+1, d_2+1)$-form field strengths or to their
dual $({\dt}_1+1, {\dt}_2+1)$-forms. To be specific,  we
assume here both of them to be of 
the  `electric'  type (generalizations to the cases
of magnetic- or mixed-type branes are then straightforward \cite{13}). Then,  
as long as $d_1\neq d_2$,  
the Lorentz \& rotational invariance along the subspaces 
fix the $(d_1, d_2)$-form potentials to
\bea
{\cal A}_1&=&{\pm} e^{X_1(z)}(dx^{\m} ){\w} (dy_1^m)\nn\\
{\cal A}_2&=&{\pm} e^{X_2(z)}(dx^{\m}){\w}(dy_2^{m' })
\label{1.3} \eea
where the ${\pm}$ signs discriminate between branes and anti-branes,  and the
differentials in the parentheses are abbreviations for the wedge products over
the indicated indices. Whenever $d_1=d_2$ and simultaneously ${\a}_1={\a}_2$, 
the above invariances fix the ansatze for ${\cA}_1$ and ${\cA}_2$ up to any
linear combinations of the above two expressions. Because in this case 
we have no mean for
distinguishing the forms in our model. Therefore, we can replace 
these two potentials by a single one written as: 
${\cA}=\pm{\cA}_1\pm{\cA}_2$.
 (Note that this property is restricted
to the orthogonal branes,  where we have: 
${\cal F}^2={\cal F}_1^2+{\cal F}_2^2$, 
so that sum of the kinetic terms of $({\cA}_1, {\cA}_2)$ 
are replaceable by that of $\cA$ only.) 

However when $d_1=d_2$ but ${\a}_1\neq{\a}_2$,  
such a replacement is not possible.
For example in Type IIB theory,  the bound state of orthogonal
D- and NS-strings excites the 2-form fields of 
the RR and NSNS sectors: ${{\cal A}}_2$ and 
${{\cal B}}_2$ respectively. Despite being orthogonal,  these 2-forms can not 
be traded for a single 2-form,  as we have: ${\a}_A=-{\a}_B=1/2$ \cite{15,23}. 
\np

{\bf The reduced Lagrangian (RL) formulation}\\
To form a reduced action describing 
the physical situation of interest, first of
all we have to specify the relevant 
degrees of freedom. In the case of 
interest these are ${\Xl}$ with $\l =1, 2$,  and
 ${\f}^{\a} =(B_0, B_1, B_2, C, {\vf})$ 
 with $\a =1, ..., 5$. Here we have unified the dilaton with the
independent metric variables into a column vector ${\f}^{\a}$ for
simplifying our later formulations. The RL for $({\f}^{\a} (z), X_{\l} (z))$  
is then found by inserting the ansatze (\ref{1.2}) and (\ref{1.3}) in the 
action (\ref{I.1}), and integrating over the relative transverse coordinates:  
$(x^{\m},y_1^m,y_2^{m'})$. This, using the formula (\ref{A.11}), gives the RL 
\be
{\cL}={\cL}_G+{\cL}_F
\label{cL}
\ee
where $\LG$ and $\LF$,  respectively the gravitational 
(including the dilaton)

and form field parts of the RL,  
have the expressions
\bea      
{\cL}_G&=&1/2e^{2G}{\Oab}{\p}{\fa}.{\p}{\fb}\nn\\
{\cL}_F&=&1/2\sum_{{\l}=1, 2}e^{2({\Xl}-{\Fl})}({\p}{\Xl})^2
\label{1.5} \eea
Here 
$\p\={\p}_a$ and $(G, {\Fl})$ are linear functions of ${\fa}$'s defined by
\bea    
G(\f )&{\;\=\;}&{\ga}{\fa}\nn\\
{\Fl}(\f )&{\;\=\;}&{\fla}{\fa}
\label{1.6} \eea
and $({\Oab}, {\ga}, {\fla})$ are matrices 
with constant entries depending on the
dimensions, 
\bea    
{\Oab}&=&2\pmatrix{{\d}({\d}-1)&{\d}{\d}_1&{\d}{\d}_2&{\dt}({\dt}+1)&0\cr
             {\d}{\d}_1 &{\d}_1({\d}_1-1)&{\d}_1{\d}_2&{\d}_1({\dt}+1)&0\cr
             {\d}{\d}_2 &{\d}_1{\d}_2 &{\d}_2({\d}_2-1)&{\d}_2({\dt}+1)&0\cr
             {\dt}({\dt}+1) &{\d}_1({\dt}+1) &{\d}_2({\dt}+1) &{\dt}({\dt}+1)&0\cr
              0&0&0&0&-1/2}\nn\\
              \nn\\
{\ga}&=&\pmatrix{{\d}/2&{\d}_1/2&{\d}_2/2&{\dt}/2&0\cr}\nn\\
f_{1{\a}}&=&\pmatrix{{\d}/2&{\d}_1/2&-{\d}_2/2&-{\dt}/2&0\cr}\nn\\
f_{2{\a}}&=&\pmatrix{{\d}/2&-{\d}_1/2&d_2/2&-{\dt}/2&0\cr}
\label{1.7} \eea
The equations of motion resulting from this RL are
\bea
{{{\d}{\cL}}\over{{\d}{\Xl}}}&=&{\p}(e^{2({\Xl}-{\Fl})}{\p}{\Xl})
-e^{2({\Xl}-{\Fl})}{({\p}{\Xl})}^2=0\nn\\
{{{\d}{\cL}}\over{{\d}{\fa}}}&=&{\p}(e^{2G}{\Oab}{\p}{\fb})
-{\ga} e^{2G}{\Obc}{\p}{\fb}.{\p}{\fc}
+\sum_{{\l}=1, 2}{\fla} e^{2({\Xl}-{\Fl})}({\p}{\Xl})^2=0
\label{1.8} \eea
Obviously a general solution for such a set of nonlinear $2^{nd}$ order
equations is not easy to find. However we will see that a class of solutions, 
corresponding
to a pair of $localized$  
distributions $\rlz$,  can be found using simple tricks. 
The suitable 
boundary conditions for these 
configurations are those of the asymptotic
flat metric, 
constant dilaton and vanishing field strengths at infinity,  i.e.
\be
{\fa}|_{z\rightarrow\infty}={\fa}_0 \;\;\;\; 
, \;\;\;\;     {\Xl}|_{z\rightarrow\infty}={\Xl}_0
\label{1.9}
\ee
In what follows we always 
take ${\fa}_0=0$, which is always possible by choosing
suitable scales of 
coordinates (for ${\a}=1, ..., 4$), and of the form fields
(for ${\a}=5$) (Note that a constant 
shift in the dilaton field can beabsorbed by scaling of the form-fields,  
as far as the Chern-Simons 
terms are not introduced in the action (\ref{I.1})).
Then of course we will not be at liberty on the definitions
of ${\Xl}_0$'s, since as will be 
seen below, they have absolute physical meanings
in terms of the mass to charge 
ratios. In particular when ${\Xl}_0=0$ we will find 
that our solution describes bound state of super p-branes.\\

{\bf Solving the equations using the constraints}\\
Although the equations (\ref{1.8}) 
seem formidable to solve, if the exponentials
can be taken (consistently) 
to be constants,  then they will be greatly simplified.
So we look for the solutions satisfying 
the constraints
\bea
&\;& G({\bff})=0      \nn\\
&\;&{\Xl}-{\Fl}({\bff})={\Xl}_0
\label{1.10} \eea
The constants on the right sides have been fixed using the 
boundary conditions
(\ref{1.9}) with ${\fa}_0=0$. 
Later we will find that these constraints have the
physical interpretations as the  
`extremality'  and the  `no-force'  conditions 
respectively. Using only the 
latter constraints in the ${\Xl}$'s \eom we
find that 
\be
{\p}^2e^{-\Xl}=0
\label{1.11}
\ee
(More precisely 
equation (\ref{1.11}) must be written as a Poisson 
equation: ${\p}^2e^{-\Xl (z)}=-\kl\rlz$. 
However, assuming that distributions 
fill only a local region of the $z$ space, eq.
(\ref{1.11}) can still be used in the outer `empty' region.) 

The general solution 
to this equation with the boundary condition as in (\ref{1.9}) is written as
\be
 e^{-{\Xl}(z)}={\k}_{\l}^{-1}{\Hl}(z)
\label{1.12}
\ee
where the constants $\kl$ and the harmonic functions ${\Hl}$ are defined as
\be
 {\kl}=e^{{\Xl}_0}
\label{1.13}
\ee
\be
{\Hl}(z)=1+{\izp}{\rl}(z' )G_{{\dt}+2}(z, z' )
\label{1.14}
\ee
$G_{{\dt}+2}(z, z' )$ is the Green's function of the $({\dt}+2)$-dimensional
(flat space) Laplacian, 
\bea                                                    
G_{{\dt}+2}(z, z' )=\left\{ \begin{array}{lll} 
{1 \over |z-z' |^{\dt}}\;\; &, &\;\;  {\dt}\neq 0 \\
-\ln|z-z' |\;\;      &, & \;\; {\dt}=0                                                        
\end{array} \right.
\label{1.15}
\eea
The masses and charges of these distributions are found in section 6
to be proportional to the integrals of corresponding
${\rl}(z)$'s, showing that these describe  `densities'  of the distributions.
Applying our constraints to the second equation in (\ref{1.8}),  we find
\be
{\Oab}{\p}^2{\fb}-{\ga}{\Obc}{\p}{\fb}.{\p}{\fc}+\sum_{{\l}=1, 2}{\kl}^2{\fla}({\p}{\Xl})^2=0
\label{1.16}
\ee
We can solve these equations for ${\fa}$'s 
using a further ansatz which is proposed
naturally by our constraints (\ref{1.10}). 
Linearity of these constraints in both
$(\fa ,\Xl)$ suggests that we take the $linear$ ansatz:
\be
{\fa}(X)=-\sum_{\l =1,2}{\xla}({\Xl}-{\Xl}_0), 
\label{1.17}
\ee
with suitable choices of the 
coefficients ${\xla}$,  we will find a simultaneous
solution of the 
constraints (\ref{1.10}) 
and the equations of motion (\ref{1.16}) for ${\fa}(z)$.
Assuming ${\Xl}$'s to be independent (which is always possible by
independent choices of the 
distributions ${\rlz}$), our constraints (\ref{1.10}) and
equations of 
motion (\ref{1.16}) are respectively 
replaced by the two sets of algebraic equations:
  \bea
&\;&  {\bg}.{\xl}=0\nn\\
&\;&  {\bfl}.{\xlp}=-{\dllp}
\label{1.18}
\eea
  and
  \bea
  {\bO}{\xl}+{\bg}({\xl}{\bO}{\xl})-{\kl}^2{\bfl}=0\nn\\
  {\xl}{\bO}{\xlp}=0 \;\;\; , \;\;\;    {\l}\neq{\lp}
  \label{1.19}
  \eea
where we have used the matrix notation for simplicity. These
 equations have a solution for $({\xaa}, {\xba})$ as,
 \bea
 {\xaa}&=&{\ka}^2(-{{\dta} \over {2{\Db}}}, 
 -{{\dta} \over {2{\Db}}}, 
 {{d_1} \over {2{\Db}}}, {{d_1} \over {2{\Db}}}, {\aa})\nn\\
 {\xba}&=&{\kb}^2(-{{\dtb} \over {2{\Db}}}, 
 {{d_2} \over {2{\Db}}}, -{{\dtb} \over {2{\Db}}}, 
 {{d_2} \over {2{\Db}}}, {\ab})
 \label{1.20}
 \eea
where $\Db\;\=\; D-2$ and $\dtl\;\=\;\Db -\dl$. 
In addition they give a set of
 `consistency conditions' , 
\bea
 {\aa}^2&=&{1 \over {{\ka}^2}}-{{d_1{\dta}} \over {2{\Db}}}\nn\\
 {\ab}^2&=&{1 \over {{\kb}^2}}-{{d_2{\dtb}} \over {2{\Db}}}\nn\\
  {\aa}{\ab}&=&{1 \over 2}({{d_1d_2} \over{\Db}}-{\d})
\label{1.21}
\eea
  which express in a different 
  notation \ft{In \cite{15} and most of the other references,  
  the $\al$'s are twice the $\al$'s in this paper.
  Also $\kl$'s are often traded for another set of parameters defined by:
  ${\D}_{\l}\equiv 4/{\kl}^2$.} the result 
  of \cite{15} for the equations governing 
  the intersections (usually called the  `intersection rules' ) of the two
  electric type branes. 
  They are derived here simply using a set of constraints.
  Generalisations of the relations (\ref{1.21}) can be obtained in a variety
  of models with various 
  scalars and form-fields corresponding to the various types
  of intersecting branes \cite{11,19}.
  We will see in section 5 that 
  equations (\ref{1.21}) have simple interpretations
  in terms of no-force conditions among the same/different types 
  of branes. For now let us 
  concentrate on the first two of them, i.e.the equation
\be
{\al}^2={1 \over{{\kl}^2}}-{{{\dl}{\dtl}} \over {2{\Db}}}
\label{al}
\ee
This relation is not specific to 
the bound states of orthogonal distributions;
even the solution to a single distribution requires such a
condition. The simple interpretation 
of (\ref{al}) is obtained by calculating the
mass $\Ml$ and charge $\Ql$ of a $(\dl -1)$-branes distribution 
(formulas (\ref{6.11}) and (\ref{6.13})),
\bea
{\Ml}&=&{{\kl}}^2{\iz}{\rlz}\nn\\
{\Ql}&=&{\kl}{\iz}{\rlz}
\label{1.22}
\eea
We observe that the physical meaning of ${\kl}$, \\
\be
{\kl}={{\Ml} \over {\Ql}}
\label{1.23}
\ee
is the mass to charge ratio. This justifies our previous statement that 
${{\Xl}}_0=\ln{{\kl}}$ can not be set by hand to arbitrary values. Using this 
ratio in (\ref{al}),  we find a relation between the three  `charges' 
\ft{These  `charges'  include the gravitational,  dilatonic and form-field
charges,  which are proportional to $\Ml$,  $\al \Ml$ and $\Ql$ respectively.
Compare to \cite{6} for a similar 
discussion in the context of black hole solutions.}
of a $(\dl -1)$-brane: 
\be
{\Ql}^2={\al}^2{{\Ml}}^2+{{\dl}{\dtl} \over 2{\Db}}{{\Ml}}^2
\label{1.24} \ee
This relation will become physically 
meaningful if we recall that the long range forces between
two parallel identical $({\dl}-1)$-branes  have the forms
\be
F_G\sim -{{{\Ml}}^2 \over r^{{\dtl}+1}}\;\;\;   ,  \;\;\;
F_D\sim -{{\al}^2{{\Ml}}^2 \over r^{{\dtl}+1}}\;\;\;   ,  \;\;\;
F_F\sim +{{{\Ql}^2} \over {r^{{\dtl}+1}}}
\label{1.25}
\ee
where$F_G$, $F_D$ and $F_F$ represent 
the gravitational, dilatonic and $({\dl}-1)$-form
forces and $r$ is the transverse distance between two branes.
In this way the equation (\ref{1.24}) 
is nothing but the simple statement that
the gravitational and dilatonic attractions of two $({\dl}-1)$-branes
 must be cancelled against their $({\dl}-1)$-form repulsion, to allow 
 stable bound state. 
 Alternatively we might consider a single $({\dl}-1)$-brane instead of
 a pair of them 
 (i.e. ${\rlz}\sim{{\d}}^{{\dtl}+1}(z)$). 
 In that case again we would recover equation (\ref{1.24}),
  but of course with a different interpretation 
  as the balance between the various long-range forces
 between different parts of a single brane. In other words,  if a brane has
 enough  `charge' ,  its form-field 
 repulsion will prevent its collapsing due to the 
 gravitational/dilatonic  
 `self attractions'  (when it has been slightly  `bended' ).
 
  For the third relation in 
  (\ref{1.21}), a similar argument must yield an interpretation
  in terms of the pair-wise no-force conditions of the intersecting branes
  (see section 5 and the discussion of \cite{23}).\\
  An interesting special case of the above 
  solution concerns the bound states of
  super p-branes. In such a case 
  the unbroken 1/2 supersymmetry of each individual
  $({\dl}-1)$-brane causes it to 
  have the minimum mass:${\Ml}=|{\Ql}|$ (i.e. ${\kl}={\pm} 1$).
  Such
  branes are of special interest,  
  because in addition to being classically stable
  against the long-range forces 
  (which is guaranteed by (\ref{1.24})),  they are also
  stable against loop corrections 
  and Hawking radiations \cite{53,54}. Strictly speaking such
  (supersymmetric) objects occur (only) in supersymmetric theories like the
  supergravity and superstring 
  theories. So not surprisingly supersymmetry of the theory requires
  special relations among its parameters so as to render it compatible with 
  the existence of such extended objects. For the model at hand a relation
  of this kind,  restricting the couplings and dimensions,  is obtained by
  putting ${{\kl}}^2=1$ in equation (\ref{al}). Thus in agreement
  with \cite{13},  we find that
  \be
  {\al}^2=1-{{\dl\dtl}\over{2\Db}}
  \label{1.26} \ee
  This equation reproduces 
  (up to a sign) the dilaton couplings to the form-fields in all
  $D=10$ supergravity theories 
  with one dilaton field. In $D=11$ supergravity
  which has no dilaton field (${\al}=0$), 
  hence we find that only the dual 2 and 5-branes
  (corresponding to 3 and 6-form potentials) 
  are possible as supersymmetric extended objects.\\
   Returning to our double brane system, from (\ref{1.21}) we obtain,
   \be
   (2\Db -d_1\dta )(2\Db -d_2\dtb )=(\d\Db -d_1d_2)^2
   \label{1.27} \ee
   We shall refer to this Diophantine equation as the
   `intersection rule' in this paper, since it specifies the suitable
  dimension of the intersection subspace for marginal configurations.
  In Appendix B methods for classifying the solutions of (\ref{1.27})
  for arbitrary $D$ are proposed and the possible solutions for $D=4,6,10,11$
  are indicated. It must be 
  noted however that there exist also BPS non-marginal 
  configurations \cite{38,18}, which do not obey 
  the intersection rule (\ref{1.27}). 
  As has been indicated in \cite{38}, the 
  non-marginal solutions
  in $D=10$ can be constructed 
  from the marginal ones using 
  T- and S-duality transformations.
  For theories in $D<10$ and $D=11$ 
  this can be done using the methods of dimensional reduction
  and oxidation \cite{20}. 
  However a general theory of such solutions for arbitrary
  spacetime dimensions is presently absent.\\

   {\bf Summary of the solutions for orthogonal $d_1\cap d_2={\d}$ system}\\
   The solutions for the metric and 
   dilaton are found from (\ref{1.17}), referring
   to the definitions of the variables 
   and values of the parameters in (\ref{1.20}).
   Also the form-potentials are already 
   known using their expressions by (\ref{1.3}) and
   (\ref{1.12}). The results for the general case (arbitrary ${\kl}$) are
   \bea
   &&ds^2=H_1^{{\ka}^2d_1/\Db}H_2^{{\kb}^2
   d_2/\Db}\{H_1^{-{\ka}^2}
   H_2^{-{\kb}^2}\dx\dxp + 
   H_1^{-{\ka}^2}\dy\dy +H_2^{-{\kb}^2}\dyp\dyp +\dz\dz\}\nn\\
   &&e^{\vf}=H_1^{{\ka}^2\aa}H_2^{{\kb}^2\ab}\nn\\
   &&{\cA}_1=\ka H_1^{-1}(\dx )\w (\dy ) \nn\\
   &&{\cA}_2=\kb H_2^{-1}(\dx )\w (\dyp )
   \label{1.28} \eea
   Therefore we see that the solutions of the reduced theory, satisfying   
    suitable constraints, are in 
    complete agreement with those of \cite{15,27}
    obtained by solving  `directly'  
    the field equations of the original action (\ref{I.1}).
   As pointed in the introduction, 
   this feature is a general property of the marginal
   solutions satisfying the $\cL =0$ condition, which is seen to be manifest
   for the above solution.\\
  
  \section{Generalization to the multiply intersecting branes }
  \setcounter{equation}{0}
  To illustrate the generality and 
  simplicity provided by applying the methods 
  introduced in section 1,  we will apply them to obtain
  the solution for $N$ (orthogonal) intersecting branes. The steps will be 
  exactly parallel to those of section 1. 
  We will use, for simplicity,  of a model with one
  dilaton field ${\vf}$ and $N$ 
  form-potentials ${\cA}_{\l}$ of degrees ${\dl}$
  with ${\l}=1, ..., N$,  
  which are coupled to $N$ orthogonal branes of electric
  type with the respective 
  world-volumes dimensions. Again the whole of the branes
  are assumed to have uniform distributions 
  within a $d$-dimensional world-volume and
  arbitrary distributions transverse 
  to it. Clearly the (minimum possible) dimension of 
  this wold-volume $(d)$ is 
  \be
  d=\sum _{\l}d_{\l}-\sum _{{\l}<{\l}' }d_{{\l}{\l}' } 
  +\sum _{{\l}<{\lp}<{\ls}}d_{{\l}{\lp}{\ls}}-...
  \label{2.1}
  \ee
  where ${\l}, {\lp}, ...=1, ..., N$,  
  and $d_{\l\lp}, d_{{\l}{\lp}{\ls}}$'s, etc. are the dimensions of  
  double, triple,  etc. intersections 
  respectively. We may assume that the time direction is included
  in all the ${\dl}$'s so that the configuration is static,  though it is not 
  an essential restriction. 
  We will use the Cartesian coordinates $(x^i)$ and $(z^a)$ for
  the homogeneous  subspace and the
  subspace transverse to it respectively.\\

  {\bf The ansatze for $ds^2$ and ${\cA}_{\l}$'s}\\
  We choose a set of 
  world-volume coordinates $(x^i)$,  whose directions span the
  various constituent branes'  
  world-volumes. In these coordinates we can write
  \bea
  ds^2&=&e^{2B_i(z)}\eta_{ij}dx^idx^j+e^{2C(z)}dz^adz^a\nn\\
  {\cA}_{\l}&=&{\pm} e^{\Xlz}{\W}_{i{\in}{\dl}}dx^i 
  \label{2.2}
  \eea
  where ${\eta}_{ij}=diag(-1, +1, ..., +1)$  ;  $i, j=0, ..., d-1$.\\
  
  {\bf The RL and all that...}\\
   The relevant degrees of freedom 
   are ${\fa}\;\=\; (B_i, C, {\vf})$ and ${\Xl}$. The
   corresponding RL is written as
   \be
   {\cL}=1/2e^{2G({\bff})}{\Oab}{\p}{\fa}.{\p}{\fb}+
   1/2\sum_{{\l}=1}^{N}e^{2({\Xl}-{\Fl}({\bff}))}({\p}{\Xl})^2
    \label{2.3}
    \ee
   where as in section 1 we have defined $(N+1)$ 
   linear functions of $({\fa})$ as
   \bea
   G({\bff})&=&{\ga}{\fa}\nn\\
   {\Fl} ({\f})&=&{\fla}{\fa} \;\; , 
   \label{2.4}
   \eea
   and the $(d+2)$-dimensional constant matrices as
   \bea
   {\Oab}&=&2\pmatrix{0&1&...&1&{\dt}+1&0\cr
                  1&0&...&1&{\dt}+1&0\cr
                  ...&...&...&...&...&...\cr
                  1&1&...&0&{\dt}+1&0\cr
                  {\dt}+1&{\dt}+1&...&{\dt}+1&{\dt}({\dt}+1)\cr
                  0&0&...&0&0&-1/2} \nn\\
                  \nn\\
   {\ga}&=&1/2(1, ..., 1, {\dt}, 0) \nn\\
   {\fla}&=&1/2({\pm}1, ..., {\pm}1, -{\dt}, -2{\al})=
   1/2({\ve}_{{\l}i}, -{\dt}, -2{\al})\;\; .
   \label{2.5}
   \eea
The symbol ${\ve}_{{\l}i}$ is
\bea                                                         
{\ve}_{{\l}i} = \left\{ \begin{array}{lll} 
 +1\;\;  &, & \;\; i\in{\dl} \\ 
 -1 \;\; &, & \;\; i\notin{\dl}                                                           
 \end{array} \right.
 \label{2.6}
 \eea
It is important to note that as far as we have not included any extra term
(e.g. Chern-Simons terms) in our 
action (\ref{I.1}),  the general form of the corresponding
RL for orthogonally intersecting branes
always resembles that of (2.3). The only difference of such models
appears in the definitions of the variables ${\fa}$ and of the constants
$({\fla}, {\ga}, {\Oab})$.
Hence a marginal orthogonal solution in every case
is obtained,  by following the procedure introduced in the
previous section. In particular one obtains the same equations
of motion,  constraints,  harmonicity condition for $e^{-{\Xl}}$'s,  
ansatz for ${\fa}(X)$, 
and consistency equations as the equations 
(\ref{1.8}),  (\ref{1.10}),  (\ref{1.12}), 
(\ref{1.17}) and (\ref{1.18}) \& (\ref{1.19}) 
respectively (but now with ${\l}=1, ..., N$).
The final link in this chain 
always is to solve the set of algebraic equations
(\ref{1.18}) \& (\ref{1.19}) 
simultaneously for finding (if exists) the consistent values
of our ${\xla}$'s. Of course the 
existence of such a consistent solution, often will
be accompanied by a set of 
relations between the parameters of the model itself,
 which as we have seen, lead to physical information about the
 underlying theory. The solution to
 the equations (\ref{1.18}) \& (\ref{1.19}) 
 together with the necessary consistency
 conditions are found by simple algebraic manipulations to be,
 \be
 {\xl}={\kl}^2{\bO}^{-1}({\bg}+{\bfl})
 \label{2.7}
 \ee
 \bea
&&{\bg}{\bO}^{-1}({\bg}+{\bfl})=0  \nn\\
&&{\bfl}{\bO}^{-1}{\bflp}={\bg}{\bO}^{-1}{\bg}-{{\dllp} \over {\kl}{\klp}}
 \label{2.8}
 \eea
 In particular for the model at hand, 
 replacing $({\bO}, {\bg}, \bfl)$ with those of (\ref{2.5})
 one obtains from (\ref{2.7}),
 \be
 {\xla}={\kl}^2(-{\2}{\d}_{{\l}i}+
 {{\dl} \over 2{\Db}},\; {{\dl} \over 2{\Db}}
 ,\; {\al})
 \label{2.9}
 \ee
 where we have defined
 \bea                                                                 
 {\d}_{{\l}i}=\left\{ \begin{array}{lll}                                         
 1 \;\; &, &\;\;  i\in{\dl}    \\
 0 \;\; &, &\;\;  i\notin{\dl}
 \end{array} \right.
 \label{2.10}
 \eea
 The first consistency condition 
 in (\ref{2.8}) becomes identity (!) while the
 second gives
 \be
 {\al}{\alp}-{{\dllp}\over {{\kl}{\klp}}}
 ={\2}\left ({{{\dl}{\dlp}} \over {\Db}}-d_{\l\lp}\right )
 \label{2.11}
 \ee
 This is a closed form of the three equations in (\ref{1.21})
  for each pair of $({\dl}-1, {\dlp}-1)$ branes intersecting over
  $(d_{{\l}{\lp}}-1)$ dimensions. 
  This equation has the simple interpretation that:\\
  for $N$ branes to be marginally bounded,  
  the various  `two-body'  forces among 
  them must vanish separately. If for example $d_{{\l}{\lp}{\ls}}$
 had been appeared in (\ref{2.11}),  
 we should have interpreted it as the vanishing
 of the  `three-body'  forces which 
 obviously was weaker than the above condition.
 The final expressions for the solutions are written as
 \bea
 ds^2&=&\prod_{{\l}=1}^N{\Hl}^{{\kl}^2{\dl}
 /{\Db}}\{(\prod_{{\l}=1}^N{\Hl}^{-{\kl}^2{\d}_{{\l}i}})
 {\eij}dx^idx^j+dz^adz^a\}\nn\\
 e^{\vf}&=&\prod_{{\l}=1}^N{\Hl}^{{\kl}^2{\al}}   \nn\\
 {\cA}_{\l}&=&{\kl}{\Hl}^{-1}{\W}_{i\in{\dl}}dx^i
 \label{2.12}
 \eea
 where $\Hl$'s are $N$ independent harmonic functions. This reproduces the 
 result of \cite{15} in a simple way,  thereby gives another proof 
 of the  `harmonic functions rule'  \cite{26}. Again ${\kl}$'s measure the 
 mass to charge ratios. 
 So when ${\kl}={\pm}1$ we obtain the solution for bound  
 state of super \pbs . In this case 
 equation (\ref{1.27}) generalizes to a pair-wise
 intersection rule of the form 
\be
(2{\Db}-{\dl}{\dtl})(2{\Db}-{\dlp}{\dtlp})
=(d_{{\l}{\lp}}{\Db}-{\dl}{\dlp})^2 \;\;\;\; , \;\;\;\;        
({\l}\neq{\lp})
\label{2.13}
\ee
a relation restricting different dimensions involved.\\

\section {General framework for (marginal) multiply intersecting branes}
\setcounter{equation}{0}
Before proceeding to discuss the case of branes at angles we need to somehow
develop the tools of the previous 
sections to include the more general cases of
marginal bound states of p-branes.
The key point about the corresponding solutions is
that they can be described in 
terms of a set of $independent$ harmonic functions, 
in one-to-one correspondence to the transverse distributions of branes.
(by definition each distribution consists of only the same type parallel
branes). In order to give a general 
formulation,  we consider a spacetime of
arbitrary dimension $D$ 
containing a homogeneous (but in general anisotropic
) time-like surface of dimension 
$d\leq D-3$. This subspace can be the world-volume
of various types of $p$-branes 
with arbitrary relative orientations (and/or velocities)
and separations,  which are 
filled (and/or moving) uniformly within a $(d-1)$-dimensional
space-like volume. Such a spacetime admits a set of Killing vectors
which generate the symmetry group:
$R^d\times SO({\dt}+2)$,  where the 
two factors represent the translational and
rotational invariances along the 
distribution space and the transverse space 
respectively. We can decentralize arbitrarily the positions
of different uniform distributions along the transverse space to obtain 
non-uniform distributions. 
This will break the $SO({\dt}+2)$ rotational symmetry, 
but not the $R^d$ translational invariance. So as in section 2,  we can
introduce two sets of Cartesian 
coordinates $(x^i)$ and $(z^a)$,  along the distribution and
the transverse space respectively. In 
these coordinates the ansatze for the
metric and various form-potentials generally are written as
\bea  ds^2&=&{\hij} (z)dx^idx^j+e^{2C(z)}dz^adz^a \nn\\
     {\cA}^r&=&{\Ari} (z)(dx^i)                     \label{3.1} \eea
where in the last formula 
\ft{In this formula we have introduced a convention,   
according to which the set of 
indices of a $p$-form field,  as well as the wedge
products over them,  are 
indicated by a representative within a parenthesis.
In this notation,  a summation 
over all the repeated indices within a parenthesis, 
including a division by $p!$ is implied.}
${\cA}^r$ is a $(p_r+1)$-form potential,  generated
by the set of all $p_r$-branes 
of the same type,  which are assumed to carry
 `electric'  charges of the 
 corresponding form-fields. This choice makes
all the ${\cA}^r$'s to have 
non-vanishing components only within the $d$ dimensional 
world-volume ( Compare to a static charge distribution in electrodynamics).
 Generalization to systems involving 
branes with  `magnetic'  charges is  straightforward.\\ 
To construct the solution for a system of $N$ marginally bound
distributions,  we have to take 
all the field variables,  say ${\f}^A$'s,  as 
functions of $N$ independent harmonic functions: $\Hl$; $\l =1, ..., N$,
\be  {\f}^A(z)={\f}^A(H_1(z), ..., H_N(z)){\;\=\;}
{\f}^A({\bH} (z))    \label{3.4} \ee
Using this in the equations of 
motion for ${\f}^A(z)$'s,  we will obtain a new 
set of equations describing ${\f}^A(H)$'s as functions on the space of 
harmonic functions. We will use 
this technique partially throughout this
paper to solve the \eoms. A systematic approach to this viewpoint,  and
its further consequences will be 
presented in a forthcoming paper \cite{37}.\\

{\bf Generation of the suitable ansatze for ${\cA}^r$'s}\\
First of all, we can use the above principle to obtain general
ansatze for the form-potentials,  which in all the non-orthogonal cases has
a non-trivial expression. For this purpose we need the kinetic term
of ${\cA}^r$ in the RL, which is calculated, using the ansatze (\ref{3.1})
\ft{In this formula and later on,  we use the convention
introduced in (\ref{3.1}). 
So in (\ref{3.7}) a multiplication over $\hijp$'s followed
by summations over $(i)$ and $(j)$,  
and a normalization factor: $1/(p_r+1)!$ are implied.}, 
\be \LF^r=-\2 e^{2G+2{\a}_r{\vf}}({\hijp}){\p}{\Ari}.{\p}{\Arj}  
\label{3.7} \ee
where $G$ is defined through
\be e^{2G}=\sqrt{-g}e^{-2C}=\sqrt{h}e^{{\dt} C}      \label{3.8}  \ee
and $h\;\=\; -det({\hij})$. 
So we find the equation of motion for ${\cA}^r$ as
\be {\p}(e^{2G+2{\a}_r{\vf}}({\hijp}){\p}{\Arj})=0         \label{3.9} \ee
which using ${\p}{\Arj}={\p}{\Hl}{\pl}{\Arj}$ and
\be      
 {\p}^2{\Hl}=0               
\label{3.10} \ee
becomes
\be  {\plp}(e^{2G+2{\a}_r{\vf}}({\hijp}){\pl}{\Arj}){\HH}=0  
\label{3.12}  \ee
The left hand side of (\ref{3.12}) 
is written as a quadratic polynomial function of the variables
${\p}{\Hl}$. Assuming independence of $\Hl$'s,  this equation
implies 
\be {\pl}(e^{2G+2{\a}_r{\vf}}{\plp}{\Arj})+{\plp}(e^{2G+2{\a}_r{\vf}}{\pl}
{\Arj})=0   \label{3.13} \ee
which upon integration yields,
\be {\Flri}\;{\;\=\;}\; e^{2G+2{\a}_r{\vf}}({\hijp}){\pl}{\Arj}=
{\clri}+{\cllp}{\Hlp}  
\label{3.14}  \ee
where ${\clri}$ and ${\cllpri}$ 
are two sets of constants with the latter having the
antisymmetry property
\be {\cllp}+{\clpl}=0   \label{3.15}  \ee
Now we argue that each of 
the ${\Ar}$'s components must be dependent only on
those subset of ${\Hl}$'s describing the charge distributions of 
the corresponding branes , 
i.e. those with $\dl =p_r+1$. In other words, 
the  form-field associated to the charges 
of a given type  must be independent of the    
charges of the other types.
This is a consequence of the 
fact that the equation of motion for ${\Ar}$, 
i.e. $d*(e^{2{\a}_r\vf}{\cF}^r)=*{\cJ}^r$ 
contains $only$ the currents of the
(corresponding) $p_r$-branes as the source term. 
This feature,  according to (\ref{3.14}), 
corresponds to the existence of a consistent solution with the properties
\bea 
for\;\; \dl\neq p_r+1\;\;\;\left\{\begin{array}{lll}
&&{\pl}{\Ari}=0\\         
&&{\clri}={\cllp}=0
\end{array} \right.
\label{3.16}
\eea
For ${\dl}=p_r+1$ we see from (\ref{3.15}) 
and (\ref{3.16}) that $\cllp$ in (\ref{3.14}) only mixes
${\Hl}$'s of the corresponding ${\Ar}$ sector. 
In fact consistency requires that all ${\cllp}$'s
in each sector to be vanishing. 
So we can write (\ref{3.14}),  after inverting it
for ${\pl}{\Ari}$,  as
\be  {\pl}{\Ari}=e^{-{\gg}}{\clrj}{\h}               
\label{3.18}  \ee
This gives a general electric ansatz for ${\Fr}$ in terms of the dilaton
and the metric components. This is simply achieved by taking the curl of
${\cA}^r$ in (\ref{3.1}) using (\ref{3.10}) and (\ref{3.18}),  the result is
\be {\cF}^r={\ori}{\w} (dx^i)  \label{3.19}  \ee
where ${{\ori}}$'s are a set of 1-forms defined by
\be {\ori}\;{\;\=\;}\;d{\Ari}=e^{-\gg}{\clri}{\h}d\Hl  \label{ori}  \ee
If we know the $\Hl$-dependences 
of $({\vf}, {\hij}, C)$,  then we can integrate
(\ref{3.18}) to express ${\Ari}$'s 
(and so ${\Ar}$'s) as explicit functions of $({\Hl})$.
Of course this is possible only 
whenever certain integrability conditions are
satisfied. These conditions 
are written simply as the Bianchi identities:
 $d{\cF}^r=0$ which are equivalent to
\be d{\ori}=0            \label{3.20}  \ee
Equation (\ref{3.20}) leads to a set of first
order partial differential 
equations 
for the unknown functions $({\vf}(H), C(H), {\hij}(H))$. Since these
functions must also satisfy the corresponding \eoms ,
 the above construction (in principle) determines the possible
 relations among ${\clri}$'s. This in turn 
 specifies the allowed set of angles between
 the various $p$-branes, since 
 as we will see shortly,  all these constants have
 expressions in terms of angles.\\

{\bf Some special cases}\\
 Before proceeding further,  it worths to examine here the above result
 for some well known marginal BPS solutions. For simplicity we will consider
 solutions with only super $p$-branes, i.e. $\kl =1$.\\
 1) $Single$ $(d-1)$-$brane$ $(N=1, n=1)$\\
In the notations of sections 1 and 2 we have the solutions
\be {\hij}=H^{{\b}-1}{\eij}\;\;\; , \;\;\; 
e^{2C}=H^{\b}\;\;\; , \;\;\; e^{\vf}=H^{\a}\;\;\; , \;\;\;  
A_{(i)}={\e}_{(i)}H^{-1}  \label{3.21}  \ee
where ${\b}{\;\=\;}d/{\Db}$. From these we find, $G=0$, and 
for $\Flri$ (dropping the $r$ and $\l$ indices) becomes
\be F^{(i)}=-{\e}^{(i)}H^{2{\a}^2-2+d{\dt} /2{\Db}}       \label{3.22}  \ee
So equation (\ref{3.14}) is satisfied, provided we have
\be  {\a}^2=1-d{\dt} /2{\Db}           \label{3.23}  \ee
which proves (\ref{1.22}) again. Under this condition we have
\be   {\cli}=-{\e}^{(i)}\;\;\;\;  ,  \;\;\;\; {\cllp}=0   
\label{3.24}   \ee
2) $Two$ $orthogonal$ $branes$: $d_1\cap d_2={\d}\;\;\;\;\; (N=2, n=2)$\\
Using the solutions (\ref{1.28}) we have
\bea
&\;&{\hij}=\pmatrix{H_1^{{\b}_1-1}
H_2^{{\b}_2-1}{\emn}&0&0\cr0&H_1^{{\b}_1-1}
H_2^{{\b}_2}{\dmn}&0\cr 0&0&H_1^{{\b}_1}
H_2^{{\b}_2-1}{\dmpnp}}\nn\\
&\;&e^{2C}=H_1^{{\b}_1}H_2^{{\b}_2}\;\;\;\;\;\;\;\;\;\;\;\;\;\;   
,  \;\;\;  e^{{\vf}}=H_1^{{\a}_1}H_2^{{\a}_2} \nn \\
&\;&A^1_{({\m})(m)}={H_1}^{-1}{\e}_{({\m})}{\e}_{(m)}\;\;\;   
,  \;\;\;  A^2_{({\m})(m' )}={H_2}^{-1}{\e}_{({\m})}{\e}_{(m' )}
\label{3.25}
\eea
where ${\b}_{\l}={\dl}/{\Db}$ 
and ${\e}_{(\m)}$,  ${\e}_{(m)}$ 
and ${\e}_{(m' )}$ are the alternating
symbols corresponding to the 
volume forms of the subspaces: $x^{\m}$ ,  $y_1^m$ and $y_2^{m' }$
respectively. All other components of the form-potentials are vanishing.
Again we find $G=0$,  and the non-vanishing ${\Flri}$'s become,
\bea  F_{\l}^{1({\m})(m)}&
=&-{\d}_{{\l} 1}{\e}^{({\m})}{\e}^{(m)}H_1^{2{\aa}^2-2+
d_1{\dta}/{\Db}} H_2^{2{\aa}{\ab}+{\d}-d_1d_2/\Db}\nn\\
      F_{\l}^{2({\m})(m' )}&
      =&-{\d}_{{\l} 2}{\e}^{({\m})}{\e}^{(m' )}
      H_2^{2{\ab}^2-2+d_2{\dtb}/{\Db}} H_1^{2{\aa}{\ab}+{\d}-d_1d_2/\Db}
\label{3.26}
\eea
The only way for these to satisfy (\ref{3.14}),  is that all the
exponents in (\ref{3.26}) vanish. This gives
\bea
         &&{\aa}^2=1-{{d_1{\dta}}\over{2{\Db}}}\nn\\
         &&{\ab}^2=1-{{d_2{\dtb}}\over{2{\Db}}}\nn\\
         &&{\aa}{\ab}={\2}\left ({{d_1d_2}\over{\Db}}-{\d}\right )
\label{3.27}
\eea
which are the same as the consistency 
conditions (\ref{1.21}) (with ${\kl}={\pm}1$).
Under these conditions we see,  in agreement with (\ref{3.14}),  that:
${\Flri}={\clri}=const$.  The non-vanishing ${\clri}$'s are
\bea  &&c_1^{1({\m})(m)}=-{\e}^{({\m})}{\e}^{(m)}\nn\\
      &&c_2^{1({\m})(m' )}=-{\e}^{({\m})}{\e}^{(m' )}
\label{3.28}
\eea

3) $Type$ $IIA$ $(2, 2)-branes$ $at$ $angles$ $(N=2, n=1)$\\
The solution to the system $3\cap 3=1$ at 
two $SU(2)$ angles in Type IIA theory, 
and its generalization for $n$ 
number of such 2-branes, was originally 
conjectured by Myers $et.al.$ in \cite{22} 
and later derived in \cite{17}. This is specially an interesting
example for comparison with the 
results of this paper. Both of the above
references have made use of an 
$asymptoticly$ $orthogonal$ Cartesian coordinate system
$(t, x_1, x_2, x_3, x_4)$ for the 
world-volume directions. In these coordinates one of the
2-branes lies on the $(x_1, x_2)$ plane while the other
is located on the plane $(x_1' , x_2' )$,  obtained from the first one
by two $SU(2)$ rotations with the same angle ${\th}$,
\bea  &&x' _1+ix' _3=e^{i{\th}}(x_1+ix_3)\nn\\
      &&x' _2+ix' _4=e^{-i{\th}}(x_2+ix_4)
\label{3.29}
\eea
In these coordinates their solutions (in the Einstein frame) can be put into
the forms
\bea
ds^2=&&E^{-5/8}[-dt^2+(1+h_1+{\cbt} h_2)(dx_1^2+dx_2^2)
+(1+{\sbt} h_2)(dx_3^2+dx_4^2)\nn\\
       &&+2{\ct}{\st} h_2(-dx_1dx_3+dx_2dx_4)]+E^{3/8}dz^adz^a\nn\\
{\cA}=&&E^{-1}dt{\w}[({\cbt} h_1+
h_2+{\sbt} h_1h_2)dx_1{\w} dx_2-{\sbt} h_1(1+h_2)dx_3{\w} dx_4\nn\\
      &&+{\ct} {\st} h_1(dx_1{\w} dx_3+dx_2{\w} dx_4)]\nn\\
e^{\vf}=&&E^{1/4}
\label{3.30}
\eea
where $a=1, ..., 5$,  $h_1$ and $h_2$ are
independent harmonic functions vanishing at infinity,  and $E$ is defined by
\be   E\;{\;\=\;}\; 1+h_1+h_2+{\sbt} h_1h_2
\label{3.31}
\ee
These solutions can be put into a symmetric form,  relative to the exchange
of the two branes,  by transforming the coordinates $(x_1, x_2, x_3, x_4)$ 
to the  `branes coordinates' :
$(x_1, x_2, x_1' , x' _2)\equiv (y_1, y_2, y_3, y_4)$, which even 
at $z\ra\infty$ do not constitute an
orthogonal frame. This simple exercise gives
\bea
&&ds^2=sin^{-2}{\th} E^{-5/8}[-{\sbt}dt^2
+(1+h_1+{\cbt} h_2)(dy_1^2+dy_2^2)+
\nn\\
&&\;\;\;\;\;\;\;\;(1+h_2+{\cbt} h_1)(dy_3^2+dy_4^2)  
-2{\ct} (1+h_1+h_2)(dy_1dy_3+dy_2dy_4)]+E^{3/8}dz^adz^a\nn\\
&&{\cA}=E^{-1}dt{\w} [h_1(1+h_2)dy_1{\w} dy_2
+[h_2(1+h_1)dy_3{\w} dy_4-{\ct} h_1h_2(dy_1{\w} dy_4+dy_3{\w} dy_2)]
\nn\\
&&e^{2{\a}{\vf}}=E^{1/8}
\label{3.32}
\eea
where we have used the value of ${\a}=1/4$ for this case. 
From the above solutions 
we find $G=-\ln\st$.
We obtain the non-vanishing components of $\Fli$ as
\bea
&&F_1^{012}={1\over{\sbt}}\;\;\;\;\;\;\;\;\;\; , 
\;\;\;\;\;\;\;\;\;\; F_2^{012}={{\cbt}\over{\sbt} }\nn\\
&&F_1^{034}={{\cbt}\over{\sbt}}\;\;\;\;\;\;\;\;\;\; , 
\;\;\;\;\;\;\;\;\;\; F_2^{034}={1\over{\sbt} }\nn\\
&&F_1^{014}={{\ct}\over{\sbt}}\;\;\;\;\;\;\;\;\;\; 
, \;\;\;\;\;\;\;\;\;\; F_2^{014}={{\ct}\over{\sbt} }\nn\\
&&F_1^{032}={{\ct}\over{\sbt}}\;\;\;\;\;\;\;\;\;\; , 
\;\;\;\;\;\;\;\;\;\; F_2^{032}={{\ct}\over{\sbt} }
\label{3.34}
\eea
which are seen to be constants, confirming our equation (\ref{3.14})
with $\cllp =0$.\\

{\bf General formula for ${\clri}$}\\
The main lesson we learn from the previous examples,  is that the integration 
constants ${{\clri}}$ in the general ansatz (\ref{3.16}) are determined by
relative orientations of the constituent branes.
For this reason,  we call them as the  `structure constants'  
of the brane system.
A general formula for these constants,  can be obtained
using the boundary conditions. For localized 
distributions (with $\dt >0$), the
metric and dilaton boundary conditions are written as
\be   {\hij}(z)|_{z{\ra}\infty}={\gij}\;\;\; , 
\;\;\;  C(z)|_{z{\ra}\infty}=0\;\;\; , 
\;\;\; {\vf}(z)|_{z{\ra}\infty} ={\vf}_0=0
\label{3.35}
\ee
where ${\gij}$ is a flat Minkowski 
metric with the signature $(-, +, ..., +)$ expressed
in the (generally) non-orthogonal 
Cartesian coordinate system $(x^i)$, and for convenience
we have chosen ${\vf}_0=0$ (c.f. section 1). The other
set of boundary conditions are related to the form-potentials.
These boundary conditions arise since as $z\ra\infty$,  the gravity
and dilaton fields become so weak 
as to decouple (to first order) from all the
the form-fields. So at infinity,  
the set of form-fields propagate like an assembly of
$uncoupled$  fields (of generally different degrees) within a flat
background. These are described by the  `Maxwell's equations' :
\be  d*d{\cA}^r=*{\cJ}^r    \label{3.36}  \ee
in the flat spacetime,  with $p_r$-form sources as
\be  {\cJ}^r=\sum_{{\dl} =p_r+1}{\rlz} {\el}   \label{3.37}  \ee
Here $\el$ and ${\rlz}$ denote  respectively the $asymptotic$ volume-form and
the density function associated with the ${\l}^{th}$ 
distribution\ft{This form 
of currents is read from the $p_r$-branes $\s$-model sources,  coupled to 
the supergravity action. In the $x^i$'s basis: ${\el}\;\equiv\;{\eli} {\dxi}$  
where the constants $\eli$ for 
non-orthogonal $x^i$'s may be other than $0,\pm 1$.}. 
Solution to (\ref{3.36}) is then:
\be      {\Ari} (z)=\sum_{{\dl} =p_r+1}{\eli}{\hlz}        
\label{3.39}    \ee
where ${\hlz}$'s are harmonic functions defined by
\be       {\hlz}={\izp} {\rl} (z' )G_{{\dt}+2}(z, z' )
\label{3.40}    \ee
Note that ${\hlz}\ra 0$ as $z\rightarrow\infty$,  provided that $\rlz$
is a localized distribution.\\
Viewing equation (\ref{3.39}) as the boundary conditions for
the original problem,  we find that
\be        {\pl}{\Ari} |_{z{\ra}\infty}={\d}_{{\dl}, p_r+1} {\eli}    
\label{3.41}    \ee
Then using the boundary values (\ref{3.35}) 
and (\ref{3.41}) into the equation
(\ref{3.18}), we obtain the desired formula for ${\clri}$: 
\be        {\clri} ={\d}_{{\dl}, p_r+1}\sqrt{\g}({\gijp}){\elj}             
\label{3.42}      \ee
where $\g\= |det(\gij )|$.
An important caution for later 
applications of this formula is that it can be
used $only$ in a  `basis'  of harmonic 
functions as is defined by (\ref{3.40}).\\

{\bf Change of the harmonic functions basis}\\
For the equations describing ${\f}^A(H)$'s 
(called as the  `embedding equations' ), 
we have an obvious symmetry under 
the permutations of $\Hl$'s,  as these equations
do not involve any explicit 
dependence on $\Hl$'s. This symmetry is a special
case of a largest symmetry,  i.e.
\be      {\hat{H}}_{\l} =a_{{\l}{\k} }H_{\k} +b_{\l}\;\;\; , \;\;\; 
det(a_{{\l}{\k}})\neq 0
\label{3.43} \ee
which transforms the embedding 
equations covariantly (see (\ref{3.68})). This
originates from the fact that 
any nonsingular linear transformation on a set 
of independent harmonic functions,  
transforms them to another such set. It turns  
out that $\Hl$'s are similar to a set of Cartesian coordinates 
(called as the  `$H$-basis' ) on a kind of
 `geodesic'  surface 
 (called as the  `$H$-surface' ) in the space of field variables
\cite{37}. Symmetry of the embedding equations 
under (\ref{3.43}) is a reflection 
of the fact that,  we have not any 
preferred origin and direction for choosing 
the $H$-basis on the $H$-surface. The 
transformation property of $\clri$'s
under (\ref{3.43}),  can be read from 
their definition by (\ref{3.14}) to be
\be       \ch_{\l}^{r(i)}=(a^{-1})_{{\k}{\l} }c_{\k}^{r(i)}          
\label{3.44}   \ee
In section 4 we will find that the solution 
for a system of two \pbs at angles
more easily described,  in a certain basis which is related to the 
 `natural'  one (\ref{3.40}) by a suitable 
 transformation (see (\ref{4.53})).\\

{\bf The reduced Lagrangian}\\
The gravitational and form-field 
parts of the total RL for $(\hij , C)$ and $\Ari$, 
are given by the equations (\ref{3.7}) 
and (\ref{A.12}) respectively. Adding the dilaton
term to them,  the total RL will read
\bea       {\cL} =&&\2 e^{2G}[\2{\trb} -\2{\tr}^2  
+2{{\dtt}} {\p} C.{\tr}\nn\\ &&+2{\dt}{{\dtt}} ({\p} C)^2
-({\p}{\vf})^2-\sum_{r=1}^ne^{2{\a}_r{\vf}} ({\hijp} ){\p}{\Ari}.{\p}{\Arj}]
\label{3.45}
\eea
where ${\p}\equiv{\p}_a$, and $G$ is defined as in (\ref{3.8}). 
Before discussing the
equations of motion,  some simplifications is in order:\\
$a$) As in sections 1 and 2 we take the benefit of the constraint: $G=G_0$,  
which means that $C$ must be replaced by
\be     C=-{1\over {4{\dt}}}(\ln |{\bh} |-\ln |{\bga} |)          
\label{3.46}    \ee
where we have set $G_0=1/2\ln\sqrt{|{\bga}|}$ 
using the boundary conditions (\ref{3.35}).
Then by eliminating $C$ 
between (\ref{3.45}) and (\ref{3.46}),  the simplified RL
describing $(\hij , \vf, \Ari )$ becomes
\be    {\cL} =-1/4[{\dt}^{-1}{\trb} 
+{\tr}^2]-1/2({\p}{\vf})^2
-1/2\sum_{r=1}^ne^{2{\a}_r{\vf}} ({\hijp}){\p}{\Ari}.{\p}{\Arj}
\label{3.47}   \ee
Demanding solutions to be on the  `surface' : $G=G_0$ requires a consistency
condition. This is obtained by a 
simple elimination of $C$ from its \eom using 
(\ref{3.45}) and (\ref{3.46}). For $\dt\neq 0$ this yields
\be      {\cL}=0
\label{3.48}   \ee
which will be interpreted as the  `marginality'  condition.\\
$b$) As we have seen in sections 1 and 2,  the number of actual
metric variables for a given system,  is often less than
that of the ${\hij}$'s. Denoting these variables together
with the dilaton collectively by ${\f}^{\a}$                         
($\a =1, ..., M+1$ where ${\f}^{M+1}\;\=\;\vf$), 
the gravitational (including the dilaton) 
part of (\ref{3.47}) will take a simple form as  
\be        \LG =1/2{\Oab} ({\bff} ){\p}{\fa}.{\p}{\fb}
\label{3.50}\ee
Here $\Oab$'s are known functions of ${\f}^{\a}$'s. This Lagrangian
looks like a kind of (pseudo-) Riemannian 
metric on the space of ${\f}^{\a}$'s, 
with $\Oab$ as its metric tensor \cite{37}.\\
$c$) Let us consider, for a 
moment, ${\Hl}$'s as the set of independent variables required to
describe all the non-vanishing 
form-fields for the configuration of interest,  
i.e. ${\Ari}={\Ari}({\Hl})$. 
So the set of all independent variables is here
 $({\fa}, {\Hl})$,  in terms of 
 which the form-field part of (\ref{3.47}) is written
as
\be   \LF =-\2{\Ul} ({\bff} , {\bH} ){\p}{\Hl} .{\p}{\Hlp}
\label{3.51} \ee
with 
\be    \Ul (\f , H)\;\=\; \sum_{r=1}^ne^{2{\a}_r{\vf}} 
({\hijp} ){\pl}{\Ari}{\plp}{\Arj}
\label{Ul} \ee
In the last expression the (known) $({\fa})$-dependence of $\Ul$
comes through $({\vf}, {\hij})$,  
while its (unknown) $({\Hl})$-dependence comes through
${\Ari}$.\\ 
Using these points,  
the complete Lagrangian describing $({\fa}, {\Hl})$ becomes
\be        {\cL} =1/2{\Oabf} {\p}{\fa} 
.{\p}{\fb} -1/2{\Ulf} {\p}{\Hl} .{\p}{\Hlp}
\label{3.52}\ee
The \eom for ${\Hl}$ from this RL is written as
\be        {\Ul} {\p}^2{\Hlp} +{\p}{\Ul} 
.{\p}{\Hlp} -1/2U_{{\lp}{\ls}, {\l}}\p{H_{\lp}}.\p{H_{\ls}} =0
\label{3.53}\ee
where the symbol  `$, {\l}$'  in the 
subscript represents the partial derivatives 
with respect to the explicit
$\Hl$-dependences,  wheres ${\pl}$ will represent the total 
${\Hl}$-derivatives. Now we enter the main 
three assumptions characterizing the set of
marginal solutions as:\\
1) ${\p}^2{\Hl}=0$ outside the region of branes distributions.\\
2) ${\fa}$'s can also be expressed as functions of ${\Hl}$'s.\\
3) ${\Hl}$'s constitute a set of $independent$ harmonic functions.\\
Using the above ingredients in the last equation,  we obtain
\be       {\pls} U_{{\l}{\lp} }+{\plp} 
U_{{\l}{\ls} }-U_{{\lp}{\ls} , {\l} }=0
\label{3.54}\ee
It is clear from (\ref{Ul}) that,  the functional
dependences of ${\Ul}$'s on $({\Hl})$ 
will not be known,  unless we specify those
of ${\Ari}$'s. We do not know these 
functions presently,  but rather we have (from 
(\ref{3.18})) the expressions 
for their derivatives ${\pl}{\Ari}$ as functions 
of $({\fa})$, which using $G=G_0$,  are written as,
\be       {\pl}{\Ari} =e^{-2{\a}_r{\vf}}{\clrj} {\h}
\label{3.55}\ee
Using this result in (\ref{3.51}),  
an explicit solution for ${\Ul}$ on the $H$-surface
is find to be
\be       {\Ul} ({\bff} ({\bH} ), {\bH} )={\ul} ({\bff} ({\bH} ))
\label{3.56}\ee
where ${\ul}$'s are given by 
\be        {\ul} ({\bff} )\;{\;\=\;}\; 
\sum_{r=1}^ne^{-2{\a}_r{\vf}}{\clri} c_{\lp}^{r(j)} {\h}
\label{3.57}\ee
These functions are written as the sum of $n$ homogeneous polynomials
in the variables $({\hij})$, 
of degrees $(p_1+1), ..., (p_n+1)$,  with the coefficients
which are combinations of the 
structure constants. We will use, in section 4,
these polynomial functions,  
to analyze the field equations of two \pbs at angles.\\
In Appendix D it is shown that the 
above expression is in fact the solution
of the equation (\ref{3.54}) 
for $\Ul$ on the $H$-surface. An important result of this
analysis is that all the functions ${\Ul}$ 
(and so ${\ul}$) on the $H$-surface can be
derived from a kind of  `potential'  
function $U(H)$, 
\be           {\Ul} |_H={\ul} |_H={\pl}{\plp} U
\label{3.58}\ee
which is unique up to any linear function of $({\Hl})$.
(By this result,  for 
solutions on the $H$-surface,  we obtain: 
$\LF =-\2{\p}^2U$. Then 
the $\cL =0$ condition gives: $\LG =\2{\p}^2U$).

This property has an important role in the proof of integrability for a
system of multiply intersecting branes \cite{37}.\\ 

{\bf Embedding equations of the $H$-surface}\\
Having expressed every thing in terms of $({\fa})$,  
it now suffices to set the
equations determining the  `embedding functions'  
${\fa}={\fa}(H)$,  to complete
the solutions to the \eoms. 
These may be found simply by starting from the ${\fa}$'s
\eom derived from (\ref{3.52}):
\be     {\Oabf} {\p}^2{\fb} 
+{\G}_{{\a}{\b}{\c}}({\bff} ){\p}{\fb} 
.{\p}{\fc} +1/2U_{{\l}{\lp} , {\a}}({\bff} , {\bH} ){\HH} =0
\label{3.63}\ee
where ${\G}_{{\a}{\b}{\c}}$ is defined in terms of ${\Oab}$ as
\be     {\G}_{{\a}{\b}{\c}}{\;\=\;}{\2} 
({\O}_{{\a}{\b} , {\c}}-{\O}_{{\b}{\c} , {\a}}+{\O}_{{\c}{\a} , {\b}})
\label{3.64}\ee
Of course the set of equations (\ref{3.63}) 
can not be solved unless we have
specified the explicit forms 
of ${\Ul}$'s as functions of $(\fa , \Hl )$,  which
is not the case at this point. 
But we can use a surprizing property of these functions,
\be       U_{{\l}{\lp} , {\a}}({\bff} , {\bH} )
=-u_{{\l}{\lp} , {\a}}({\bff} )
\label{3.65}\ee
(which gives the derivatives $U_{{\l}{\lp} , {\a}}$  as $given$
functions of $(\fa )$ on the 
$H$-surface, see Appendix D),  to rewrite (\ref{3.63}) as
\be       {\Oabf} {\p}^2{\fb} 
+{\G}_{{\a}{\b}{\c}}({\bff} ){\p}{\fb} .{\p}{\fc} 
-1/2u_{{\l}{\lp} , {\a}}({\bff}){\HH} =0
\label{3.66}\ee
which is now a well-defined equation.
Looking for solutions on the $H$-surface, 
is equivalent to using the above three 
assumptions in this equation. This leads to
\be       {\Oabf} {\pl}{\plp}{\fb} 
+{\G}_{{\a}{\b}{\c}}({\bff} ){\pl}{\fb} {\plp}{\fc} 
-1/2u_{{\l}{\lp} , {\a}}({\bff})=0
\label{3.68}\ee
This equation for the `embedding'  
of the $H$-surface, is similar to a 
a forced geodesic equation corresponding to the metric defined
by $\Oab (\bff)$. As stated earlier,  
this equation does not involve any explicit
${\Hl}$-dependences,  and is 
symmetric under the transformation (\ref{3.43})
provided ${\ul}$'s transform as
\be     {\hat u}_{{\l}{\lp}}
=(a^{-1})_{{\k}{\l}}(a^{-1})_{{\kp}{\lp}}u_{{\k}{\kp}}
\label{3.69} \ee
One can check,  using the 
definition of $\ul$,  that this is in fact the case.\\
It is instructive to note that equations (\ref{3.68}) can be put into
the form of a true geodesic equation (i.e. without force), if one
extends the configuration space 
to ${\f}^A=(\fa ,\Hl)$ and defines its metric by
\be
dS^2={\O}_{AB}d{\f}^Ad{\f}^B{\;\=\;}\Oab d\fa d\fb -\Ul d\Hl d\Hlp
\ee
Then, of course, the $H$-surface will be
a null-geodesic surface in the
space of ${\f}^A$'s.\\

{\bf Constraints}\\
The formulation described above 
provides a general framework for studying the structure
and properties of the marginal 
solutions. In particular 
it has the potential 
to give the constraints of 
sections 1 \& 2,  and 
their generalizations to the
systems of branes at angles,  
in a systematic and exclusive way. We postpone
such issues to a forthcoming 
article \cite{37}. Here we give heuristic 
derivations for the generalized 
versions of the constraints (\ref{1.10}),  for
arbitrary systems of (multiply-intersecting) 
branes at angles. The justification
for use of these constraints is 
that they have simultaneous solutions with the
\eoms . This can be verified partially,  
using the solutions in sections 
1,  2 and 4 of this paper. 
Generally the constraints for a marginal configuration, 
consist of the two following categories:\\
$a$) $Extremality$ $condition$:\\
This constraint is somewhat related 
to the  `marginality'  condition,  which states  
that the binding potential energy of 
such a system,  for arbitrary positions of its
constituents,  vanishes. In the language of fields,  this means that
the Hamiltonian density of interactions $\cH$ must vanishes everywhere. So on
a time-independent marginal solution with $\cH =-\cL$,  
we must have $\cL =0$.
Noting the $C$'s \eom as 
\be       2{{\dtt}}{\p}^2e^{2G}-1/2{\dt}{\cL}=0
\label{3.70}\ee
we see that a marginal solution satisfies
\be        {\p}^2e^{2G}=0
\label{3.71}\ee
i.e. the combination $e^{2G}$ of the metric components must be a harmonic 
function. The most natural choice is to take $G=G_0$,  or equivalently 
\be       \ln{\sqrt{|{\bh} |}}+{\dt} C=\ln{\sqrt{|{\bga} |}}
\label{3.72}\ee
It must be remembered that the ${\cL}=0$ is not a 
single equation in this formulation;
it contains a set of component equations. 
This is because,  by evaluating
the RL on the $H$-surface we obtain
\be       {\cL} =1/2[{\Oabf} {\pl}{\fa} {\plp}{\fb} -{\ulf}]{\HH}
\label{3.73}\ee
By independence of $\Hl$'s,  the ${\cL}=0$ equation implies that
\be  {\cL}_{{\l}{\lp}}\;{\;\=\;}\; {\Oabf} {\pl}{\fa} {\plp}{\fb} -{\ulf}=0
\label{3.74}\ee
which introduces a set of $\2 N(N+1)$ equations (of $1^{st}$ order),  
satisfied
by the $(M+1)$ functions $\fa (H)$. If we have enough number of constraints, 
we can combine them with these equations to solve for $\fa (H)$'s; a method
which will be used in the next section.\\
$b$) $No-force$ $conditions$:\\
There are another set of constraints, though not seemingly expressed
as the no-force conditions,  but are equivalent to the latter.
The no-force conditions corresponding to a configuration of $N$ distributions, 
are written as a 
set of $N$ constraints relating the dilaton,  metric and form-fields
components together (see section 5). 
By expressing (formally) the form-fields 
in terms of the harmonic functions,  
we can re-express these conditions
as relations between $(\fa , \Hl )$'s. 
We obtain such relations in the following
using a heuristic way of derivation: 
We consider a special case where all the charge 
distributions except the ${\l}^{th}$ one are vanishingly small, i.e.
${\r}_{\k}(z)\ra 0$ for every $\k\neq\l$. 
The ${\cA}^r$'s solution for such
a single distribution is known to be
\be     {\cA}^r=-{\d}_{{\dl} , p_r+1}{\Hl}^{-1}{\el}
\label{3.75}\ee
where ${\Hl}=1+{\hl}$ with ${\hl}$ 
as in (\ref{3.40}). The corresponding part of
the RL is calculated to be
\be        {\cL}_{F\l} =1/2e^{-2{\Fl} ({\bff} )}({\p}{\Hl}^{-1})^2
\label{3.76}\ee
where the function ${\Fl}({\bff})$ 
is implicitly defined by  
\be         e^{-2\Fl (\bff )}{\;\=\;} -e^{2(G+{\al}{\vf})}{\hp} {\eli}{\elj}
\label{3.77}\ee
It is easy to check that the ${\Hl}$'s \eom 
from (\ref{3.76}) becomes a Laplace equation, 
only if we have 
\be          {\Fl} ({\bff} )=-\ln{{\Hl}}+const.
\label{3.78}\ee
This gives the desired constraints,  
which are claimed to be valid for a general
situation.
Note that in the special 
cases where $x^i$'s are chosen to 
be the  branes' world-volume coordinates,  
$\eli$ becomes the 
alternating symbol corresponding to the coordinates 
on $(\dl -1)$-branes'  world-volume,  
and the definition (\ref{3.77}) simplifies to 
$e^{2\Fl (\bff )}=e^{2(G+\al\vf)}|det\hp|_{i, j\in\dl}$.

\section{Application to a system of branes at angles}
\setcounter{equation}{0}
An illustrative example of the  
formulation  of section 3, is provided
by considering the distributed system: $d_1\cap d_2=\d$
at non-trivial angles. 
In the branes' world-volumes coordinates (as defined in section 1)
the ansatz for the metric is written as
\bea   
ds^2=&&e^{2B_0(z)}dx^{\m} dx_{\m} +
e^{2B_1(z)}dy_1^mdy_1^m+e^{2B_2(z)}dy_2^{m' }dy_2^{m' }\nn\\&&+
     2e^{B_1(z)+B_2(z)}q(z){\gmmp}{\dy}{\dyp}+ e^{2C(z)}dz^adz^a
\label{4.1}  \eea
This differs from its orthogonal analogue,  
equation (\ref{1.2}),  by an off-diagonal
term,  which appears 
due to the non-trivial angles between the branes. 
In this expression $q$ is 
some new degree of freedom and ${{\gmmp}}$'s are
${\d}_1{\d}_2$ parameters defining the angles.
The internal (world-volume) part of the metric tensor here is written as 
\be      {\hij}=\pmatrix{
e^{2B_0}{\emn}&0&0\cr0&e^{2B_1}{\dmn}&qe^{B_1+B_2}{\g}_{mn' }\cr 
0&qe^{B_1+B_2}{\g}_{nm' }&e^{2B_2}{\dmpnp}}
\label{4.3}  \ee
Rotating the internal coordinates on the individual branes changes the 
matrix $\bga{\;\=\;}({\g}_{mm'})$ by a transformation of the form: 
$\bga\ra A\bga B^T$, where $A\in SO(\d_1)$ and $B\in SO(\d_2)$. 
However, such different $\bga$'s describe geometrically equivalent angular
situations.
The $SO({\d}_1)\times SO({\d}_2)$-rotationally
invariant parameters are the    
angles: $\{{\th}_1, ..., {\th}_m\}$, 
defined through the diagonalization of 
the matrix $\bga{\bga}^T$ (see Appendix C), whose
number is 
\be      m=Min({\d}_1, {\d}_2)=Min(d_1, d_2)-{\d}    \label{4.2}  \ee
This is smaller than the total number of $\gmm'$'s. 

{\bf The expression for $\LG$}\\
With the aid of the formulas \ref{A.12} to \ref{A.20} of the Appendix, one
obtains the RL,
\be  \LG =f(q)^{1/2}e^{2G({\bff})}\{1/2{\Ohab}(q){\p}{\fa}
.{\p}{\fb}+a(q){\sa}{\p}{\fa}.{\p}q+1/2b(q)({\p}q)^2\}
\label{4.4}  \ee
where $\fa {\;\=\;} (B_0,B_1,B_2,C,\vf )$ and $G({\bff})$ is defined by 
 (\ref{1.6}) and (\ref{1.7}). Further 
we have defined the functions of $q$ as
\bea     f(q)&{\;\=\;}&det(1-q^2{\bga}{\bga}^T)\nn\\
         a(q)&{\;\=\;}&f'(q)/2f(q)\nn\\
         b(q)&{\;\=\;}&f''(q)/2f(q)\nn\\
         {\Ohab}(q)&{\;\=\;}&{\Oab}+{\oa}{\ob}qa(q)
\label{4.5}  \eea
with a prime denoting ${\p}/{\p}q$. 
The constants $(\oa , \sa )$ in these formulas
are defined 
\bea  {\oa}&{\;\=\;}&\pmatrix{0&1&-1&0&0}\nn\\
      {\sa}&{\;\=\;}&\pmatrix{2{\d}&2{\d}_1-1&2{\d}_2-1&2({\dt}+1)&0}
\label{4.6} \eea
The matrix ${\Oab}$  
in \ref{4.5} is defined by (\ref{1.7}).
According to the above expression,  
the explicit angular dependence of $\LG$
comes totally through an even polynomial $f(q)$ of degree $2m$
, whose roots $\{q_r\}$
characterize the angles between the two
branes via  $q_r^2=1/cos^2{\th}_r$.
In the orthogonal case with ${\gmmp}=0$,  
we have $f(q)= 1$ and consequently the expression for ${\cL}_G$
returns to that of eq. (\ref{cL}) \\

{\bf Intersection rules revisited}\\
A natural question is that whether the intersection rules of the orthogonal
branes remain to be valid even for 
branes at non-trivial angles ? We argue that the main equation
determining these rules can not be 
angle-dependent and so the answer to this question
is positive. First of all,  since the 
mechanism by which we obtain a marginal
solution in this paper,  is essentially the 
same for the two cases,  we expect that any 
non-orthogonal solution of this 
type to be connected to the orthogonal solutions 
(\ref{1.28}) continously through 
the parameters $({\th}_1, ..., {\th}_m)$.
So in particular,  we should expect to require three
consistency conditions analogous 
to the equations (\ref{1.21}),  relating the couplings, 
dimensions, and possibly angles. 
But the last possibility can not be true,  as
if the counterparts of equations (\ref{1.21}) 
include angles, then either the expressions
for $({\a}_1, {\a}_2)$ or the relation 
arising between dimensions (counterpart of (\ref{1.27}))
must contain angles. Neither of 
these possibilities are physical. 
Firstly $({\a}_1, {\a}_2)$ can not be
angle-dependent,  
as they are parameters of the model itself which can not depend
on the geometry. On the other 
hand a relation between dimensions and angles implies
a quantization of angles which 
has not any physical origin. Therefore in what
follows we will assume that 
the equation (\ref{1.27}) for super-\pbs is
always applicable.\\

{\bf Derivation of the solutions for $(p, p)$-branes at angles}\\
To present an application of all
the previous general formulas,  we show here the stages of derivation   
for the solutions of two `identical' branes at angles. 
So in all the previous formulas,  we 
put $d_1=d_2\;\=\; d_0$ and 
${\a}_1={\a}_2\;\=\; {\a}_0$. In this 
case we have only a single form-field  
$\cA$ of degree $d_0$.
Applying (\ref{1.27}) to this system,  
we obtain
\bea   &&{\d}_1={\d}_2=2\nn\\
      &&{\d}=d_0-2\nn \\
      &&d=d_0+2
\label{4.8} \eea
The relative orientation in this system,  
(generally) is described in terms of
two independent angles,  say $({\th}, {\th}')$. 
A simplification occurring here   
is that,  both the subspaces of: $(y_1^m)$ 
and $(y_2^{m' })$ become 2-dimensional.
So $\gmmp$ becomes a $2\times 2$ matrix,  
which after diagonalization takes the
form
\be  {\gmmp}=\pmatrix{cos{\th}&0\cr 0&cos{\th}' }
\label{gmmp} \ee
Hence,  putting: $y_1^m=(y_1, y_2)$ and $y_2^{m' }=(y_3, y_4)$,  the metric
(\ref{4.1}) reduces to
\bea     ds^2=&&e^{2B_0(z)}dx^{\m} dx_{\m} 
+e^{2B_1(z)}(dy_1^2+dy_2^2)+e^{2B_2(z)}(dy_3^2+dy_4^2)\nn\\
     &&+2e^{B_1(z)+B_2(z)}q(z)(cos{\th}dy_1dy_3+
     cos{\th}' dy_2dy_4)+ e^{2C(z)}dz^adz^a
\label{4.10} \eea
We assume in the following the 
asymptotic flat boundary conditions (provided
$d_0<D-4$),   
\be    {\fa}|_{z\ra\infty}=0\;\;\;\;\;\;\;\; ,  
\;\;\;\;\;\;\;\;  q|_{z\ra\infty}=1
\label{4.11} \ee 
        
{\bf The $H$-basis and the structure constants}\\
We begin our analysis by computing 
the structure constants ${{\cli}}$ in the
basis ${{\hl}}$, defined by the equations (\ref{3.42}) and
(\ref{3.40}) respectively. This 
requires the asymptotic form of $\hij$,  
which by (\ref{4.10}) and (\ref{4.11}) is written as
\be   {\gij}=\pmatrix{{\emn}&0\cr 0&{\grs}}
      =\pmatrix{\emn&0&0&0&0\cr 0&1&0&{\ct}&0\cr 
      0&0&1&0&{\ct}' \cr 0&{\ct}&0&1&0\cr 0&0&{\ct}' &0&1}
\label{4.12} \ee
where the indices: $r, s=1, ..., 4$  
refer to the coordinates $(y_1, ..., y_4)$.
Then the non-vanishing structure constants ${\clm}$ 
according to (\ref{3.42}) and
(\ref{4.12}) are written as
\bea  &&{\clm}={\e}^{({\m})}{\clrs}\;\;\; ;\nn \\
     &&{\clrs}\;{\;\=\;}\;\sqrt{{\g}}{\g}^{rr' }{\g}^{ss' }{\e}_{{\l}(r' s' )}
\label{4.14} \eea 
where ${\g}\;{\;\=\;}\; det({\grs})$, and ${\e}^{(m)}$ and ${\e}_{{\l}(rs)}$ 
(with ${\l}=1, 2$ and $r, s=1, ..., 4$)
are respectively the $(p-1)-$ 
and 2-dimensional alternating symbols such that: 
${\e}^{0...(p-2)}=-1$ and ${\e}_{1(12)}={\e}_{2(34)}=+1$.
A straightforward calculation using (\ref{4.12}) and (\ref{4.14}) then gives
\bea       c_1^{12}&=&c_2^{34}={\l}\nn\\
          c_1^{34}&=&c_2^{12}={\l}{\ct}{\ct}' \nn\\
          c_1^{32}&=&c_2^{14}={\l}{\ct} \nn\\
          c_1^{14}&=&c_2^{32}={\l}{\ct}' 
\label{4.15} \eea 
where ${\l}\;{\;\=\;}\; 1/\sqrt{{\g}}=1/{\st}{\st}' $ 
and all other ${\clrs}$'s are vanishing.
Note that for ${\th}={\th}' $ ,  as is so for the (2,2)-brane solution 
 (\ref{3.32}), 
the resultant values of $\clrs$ exactly reproduce the
constant values of ${\Fli}$,  which were given by (\ref{3.34}).
We now introduce a new $H$-basis by
\be   {\Hl}=1+a_{{\l}{\k}}h_{{\k}}
\label{4.16} \ee 
with a transformation matrix $(a_{{\l}{\k}})$ as is determined below. 
The constants
${\clrs}$ in the new basis $(\Hl)$ 
then is related to the old one $(\hl)$ according to
(\ref{3.44}) by
\be   {\chlrs}=(a^{-1})_{{\k}{\l}}c_{{\k}}^{rs}
\label{4.17} \ee 
To determine the  
coefficients $(a_{{\k}{\l}})$ we demand that:\\
i) $a_{{\k}{\l}}=a_{{\l}{\k}}$ so 
that our solution to be symmetric in $(H_1, H_2)$, 
as is so for $(h_1, h_2)$.\\
ii) ${\ch}_2^{12}={\ch}_1^{34}=0$,  which 
is  necessary for the integrability
of our equations.\\
iii) For ${\th}={\th}' ={\pi} /2$ we must have ${\Hl}=1+{\hl}$.\\
These together fix $(a_{{\k}{\l}})$ as
\be            a_{{\k}{\l}}=\pmatrix{1&{\ct}{\ct}'  \cr {\ct}{\ct}' &1}
\label{4.18} \ee 
Using this matrix in (\ref{4.17}),  we obtain
\bea      {\ch}_1^{12}&=&{\ch}_2^{34}={\l}  \nn\\\
          {\ch}_1^{32}&=&{\ch}_2^{14}={\l}{\n} \nn \\
          {\ch}_1^{14}&=&{\ch}_2^{32}={\l}{\n}' 
\label{4.19} \eea 
where ${\n}\;{\;\=\;}\;{\ct}{\sbt}' /(1-{\cbt}{\cbt}' )$, 
${\n}' \;{\;\=\;}\;{\ct}' {\sbt}/(1-{\cbt}{\cbt}' )$ and as in the past
$\l\;\=\; 1/sin\th sin\th ' $.\\

{\bf Explicit expressions for $\ul$'s}\\
Knowing the values of the structure constants 
in the ${{\Hl}}$ basis, we are able to obtain the functions $\ul (\bff ,q)$ 
, also in this basis, using the formula (\ref{3.50}) which reduces to
\be     {\ul}=-e^{-2(\hat{G}-{\d}B_0+{\a}_0{\vf})}w_{{\l}{\lp}}
\label{4.20} \ee 
Here a hat is used to distinguish between the analogous quantities in the  
orthogonal and at angles cases (see (\ref{4.28}) below)
, and $w_{{\l}{\lp}}$'s
are homogeneous polynomials of degree 2 in $(h_{rs})$ defined as
\be   w_{{\l}{\lp}}\;{\;\=\;}\;{\ch}_{\l}^{rs}{\ch}_{{\lp}}^{r' s' }
h_{rr' }h_{ss' }
\label{4.21} \ee 
Using the ansatz (\ref{4.10}) for the 
metric and the values of the structure
constants as in (\ref{4.19}) we can expand the above formula to obtain
\bea   w_{11}&=&{\l}^2(h_{11}^2+{\s}h_{11}h_{33}
+2{\r}h_{11}h_{13}+\tau h_{13}^2)\nn\\
       w_{22}&=&{\l}^2(h_{33}^2+{\s}h_{11}h_{33}+
       2{\r}h_{33}h_{13}+\tau h_{13}^2)\nn\\
       w_{12}&=&{\l}^2[{\tau\over{\v}}h_{11}h_{33}
       +{\r}' h_{11}h_{13}+2{\r}' h_{33}h_{13}+\v (1+{\s}) h_{13}^2]
\label{4.22} \eea 
The constants appearing in these 
formulas are defined in terms of the angles
$({\th}, {\th}' )$ as
\bea             &&{\v}{\;\=\;}{{\ct}' \over{\ct}}\;\;\;\;\;\; ,                    \nn   \\
                 &&{\r}{\;\=\;}{\v}{\n}+{\n}' \;\;\; , 
                 \;\;\; {\r}' {\;\=\;}{\v}{\n}' +{\n} \nn  \\
                 &&{\s}{\;\=\;}{\n}^2+{\n}'^2 \;\;\; , 
                 \;\;\; {\t}{\;\=\;}2{\v}{\n}{\n}' 
\label{4.23} \eea 
where ${\n}$ and ${\n}' $ have been 
defined in terms of $(\th , \th ' )$ below (\ref{4.19}).
We now introduce the `homogeneous' variables $(q, s)$,
\be      q\;{\;\=\;}\;{{h_{13}}\over{\sqrt{h_{11}h_{33}}}}\;\;\;\;\; , 
\;\;\;\;\; s\;{\;\=\;}\;\sqrt{{{h_{33}}\over{h_{11}}}}
\label{4.24} \ee 
which are invariant under the re-scalings 
of $(h_{rs})$. Then using (\ref{4.20})
and (\ref{4.22}) we obtain an expression for ${\ul}$ as
\be  {\ul}=-{\l}^2e^{2(-\hat{G}
+{\d}B_0+B_{\l}+B_{{\lp}}-{\a}_0{\vf})}{\fl}(q, s)
\;{\;\=\;}\; -{\l}^2e^{\hat{F}_{\l}+\hat{F}_{{\lp}}}{\fl}(q, s)/f(q)
\label{4.26} \ee 
where ${\fl} (q, s)$'s are three polynomials in $(q, s)$ defined as
\bea   f_{11}(q, s)&{\;\=\;}&(1+{\s}s^2)+2{\r}sq+{\t}s^2q^2   \nn \\
       f_{22}(q, s)&{\;\=\;}&(1+{{{\s}}\over{s^2}})+2{{{\r}}\over{s}}q
       +{{\t}\over{s^2}}q^2 \nn \\
       f_{12}(q, s)&{\;\=\;}&{{\t}\over{{\v}}}+{\r}' 
       (s+{1\over{s}})q+{\v}(1+{\s})q^2
\label{4.25} \eea
In (\ref{4.26}),  $\hat{G}$ and 
$\hat{{\Fl}}$ are generalized versions of 
$G$ and ${\Fl}$ for the case at angles,  i.e.
\bea      \hat{G}({\bff}, q)&{\;\=\;}&G({\bff})+1/2\ln{\sqrt{f(q)}}\nn\\
          \hat{F}_{\l}({\bff}, q)&{\;\=\;}&{\Fl}({\bff})+1/2\ln{\sqrt{f(q)}}
\label{4.27} \eea 
The explicit expressions of the functions appearing here are
\bea      G({\bff})&=&1/2({\d}B_0+2B_1+2B_2+{\dt}C)\nn\\
          F_1({\bff})&=&1/2({\d}B_0+2B_1-2B_2-{\dt}C-2{\a}_0{\vf})\nn\\
          F_2({\bff})&=&1/2({\d}B_0-2B_1+2B_2-{\dt}C-2{\a}_0{\vf})\nn\\
          f(q)&=&(1-q^2)(1-{\v}^2q^2)
\label{4.28} \eea 
The orthogonal situation is recovered by putting: $f(q)=1\;\; ,  \;\;
{\fl} (q, s)={\dllp}$, and ${\l}=1$.\\

{\bf Constraints}\\
The two classes of the 
constraints (\ref{3.72}) and (\ref{3.78}) for this case 
take the forms 
\bea             &&\hat G({\bff}, q)=1/2\ln{\sqrt{\g}} \nn\\
                 &&\hat {\Fl}({\bff}, q)=-\ln{{\Hl}}+1/2\ln{\sqrt{\g}}
\label{4.29} \eea 
where the constants on the right of these equations
are obtained from the boundary conditions (\ref{4.11}) and 
that ${\Hl}{\ra}1$ as $z{\ra}\infty$.
These constraints relate the six 
variables $({\fa}, q)$ and the two $\Hl$'s.
Equations (\ref{4.29}),  compared to their 
orthogonal analogues in (\ref{1.10}),
have the additional term $1/2\ln{\sqrt{f(q)/\g}}$. 
This suggests that one may find a 
consistent simultaneous solution of these
and the \eoms , by taking 
\be         {\fa}=-{\xla}{\Xl}+{\ea}\ln{\sqrt{f(q)/\g}}
\label{4.31} \ee 
where ${\Xl}{\;\=\;}-\ln{{\Hl}}$ 
and $(\xla , \ea )$'s are a set of undetermined
parameters (compare to (\ref{1.17})). Replacing this in (\ref{4.29}),  
and taking $(X_1, X_2, \ln\sqrt{f})$
as linearly independent functions, one obtains nine algebraic equations for
the parameters. This consists of the equations (\ref{1.18}) 
and the following equations
\be            {\bg}.{\bet}={\bfl}.{\bet}=-1/2
\label{4.32} \ee 
where ${\bg}$ and ${\bfl}$ are defined as in (\ref{1.7}). 
To completely determine these parameters,
the consistency with the \eoms must also be required.
However, by considering the `single distribution' limits one can verify that  
the values of $\xla$'s must be 
independent of the angles and in fact they do depend
only on the dimensions of the respective \pbs . 
This means that $\xla$'s are already determined as in (\ref{1.20}) 
(with $\k_1^2=\k_2^2=1$). The values of $\ea$'s are then determined
using the $\cL =0$ condition (see below).\\

{\bf The ${\cL}_{\l\lp}=0$ equations}\\
As was stated earlier, all the field variables: $(q,\fa ,A_{({\m})rs})$
must be expressible as functions of $\Hl$'s. These  
are not independent functions;  
since by (\ref{3.18}) and (\ref{4.31}), a knowledge of $q(H)$
determines all other functions of $\Hl$'s.
To determine the former we use the ${\cL}=0$ condition, which by
(\ref{3.74}) gives  
\be  e^{2\hat{G}}[{\Ohab}{\pl}{\fa}{\plp}{\fb}+
{\sa}a({\pl}{\fa}{\plp}q+{\plp}{\fa}{\pl}q)+b{\pl}q{\plp}q]
+e^{\hat{{\Fl}}+\hat{{\Flp}}}{\fl}/f=0
\label{4.34} \ee
where ${\pl} {\;\=\;} {\p}/{\p}{\Hl}$ and 
\be      s=e^{-(B_1-B_2)}=e^{-{\oa}{\fa}}
\label{4.35} \ee 
Using (\ref{4.29}) and replacing $\Hl$ with $\Xl$, 
eq.(\ref{4.34}) becomes,
\be   [a^2({\bet}{\bOh}{\bet}+2{\bs}.{\bet})+b]
{\pld}q{\plpd}q+a[({\bet}{\bOh}{\xl}+
{\bs}.{\xl}){\plpd}q+({\bet}{\bOh}{\xlp}+{\bs}.{\xlp}){\pld}q]+
{\xl}{\bOh}{\xlp}+{\fl}/f=0
\label{4.37} \ee 
where ${\pld}{\;\=\;}{\p} /{\p}{\Xl}$. 
This can be put into a more convenient form,
\be    ({\pld}q-a_{\l}(q))({\plpd}q-a_{{\lp}}(q))=b_{{\l}{\lp}}(q, s)
\label{4.38} \ee 
where we have defined;
\bea   a_{\l}&{\;\=\;}&-{a\over c}({\bet}{\bOh}{\xl}+{\bs}.{\xl})\nn\\
      b_{{\l}{\lp}}&{\;\=\;}&-{1\over c}({\xl}{\bOh}{\xlp}+{\fl}/f) \nn\\
      c&{\;\=\;}&b+a^2({\bet}{\bOh}{\bet}+2{\bs}.{\bet})
\label{4.39} \eea
Note that in (\ref{4.39}) $f, a, b, c, a_{\l}, \hat{\bO}$ 
are only rational functions of
$q$,  while $f_{{\l}{\lp}}$ and $b_{{\l}{\lp}}$ 
are rational functions of $q$
and polynomial functions (at most of degree 2) of $s$ and $1/ s$.
By subtracting the ${\l}=1, 2$ equations in \ref{4.29}, 
we obtain 
\be               s=\sqrt{{{H_1}/{H_2}}}=e^{-1/2(X_1-X_2)}
\label{4.40} \ee 
{\bf The integrability conditions}\\
The three partial differential 
equations (\ref{4.38}) (for ${\l}, {\lp}=1, 2$ ) 
have a solution if,\\
$i$) a consistency condition and   $ii$)an integrability condition
are satisfied.\\
To see this,  consider first the two equations with ${\l}={\lp}$ .
These determine ${\pld} q$'s as 
functions of $(q, {\Xl})$,  
\be           {\pld}q=a_{\l}+{\ve}_{\l}\sqrt{b_{{\l}{\l}}} 
\label{4.41} \ee 
where ${\ve}_{\l}={\pm} 1$, upon which eq. (\ref{4.38}) requires the 
 $consistency$ condition: 
\be           b_{11}b_{22}=(b_{12})^2
\label{4.42} \ee 
The $integrability$ condition follows from:
${\pad} ({\pbd} q)={\pbd} ({\pad} q)$,  which using (\ref{4.41}) gives\\
\be       {\ve}_1\sqrt{b_{11}}(4\bar{a}' b_{22}-2\bar{a}b' _{22}+
s\dot{b}_{22})-{\ve}_2\sqrt{b_{22}}(4\bar{a}' b_{11}-2\bar{a}b' _{11}+
s\dot{b}_{11})=2(b_{11}b' _{22}-b_{22}b' _{11})
\label{4.43} \ee
where  prime means $\p /\p q$ and dot means $\p /\p s$. By
symmetry we have set $a_1=a_2{\;\=\;} \bar{a}$. 
Comparing (\ref{4.43}) to an equation in the domain of $\bf Z$ 
 of the form\ft{Mathematically
this comparison is possible, since the set of polynomials (in $(s,1/s)$)
, as that of the integers,
with ordinary addition and multiplication, constitute a `ring'.} 
\be        a\sqrt{x}+b\sqrt{y}=c 
\label{4.44} \ee 
one concludes that $b_{\l\l}(q,s)$ must be written as the square of a
polynomial $p_{\l}(q,s)$ in $s$ or $1/s$,

\be        b_{{\l}{\l}}(q, s)=[p_{\l}(q, s)]^2
\label{4.45} \ee 
This statement,  as will be seen,  leads to important implications
about the angles and other parameters of the solution. Although logically
(\ref{4.45}) is only a necessary integrability 
condition for (\ref{4.41}),  in the
case at hand it provides the sufficient 
conditions as well. Explicit expressions
of $b_{{\l}{\lp}}$'s follow from their 
definitions by (\ref{4.39}) restricted to the
case: $d_0\cap d_0=d_0-2$ via the 
equations (\ref{4.25}) and (\ref{4.28}). In this
calculation we encounter various 
combinations of the parameters $({\xla}, {\ea})$, 
all of which except one are 
found to be calculable from the equations
(\ref{1.18}), (\ref{1.19}) and (\ref{4.32}).
For later reference we list them below
\bea  {\xl}{\bO}{\xlp}=-{\d}_{{\l}{\lp}}\;\;\;\;&, 
&\;\;\;\; {\xl}{\bo}{\xlp}=1/4(2{\d}_{{\l}{\lp}}-1)\nn\\
              {\bet}{\bO}{\xl}=-1        \;\;\;\;&, 
              &\;\;\;\; {\bet}{\bo}{\xl}={\bet}{\bo}{\bet}=0 \nn\\
               {\bs}.{\xl}=1/2        \;\;\;\;&, 
               &\;\;\;\; {\bet}{\bO}{\bet}+2{\bs}.{\bet}{\;\=\;}{\k}
\label{4.46} \eea
Only the last combination $({\k})$ 
remains unknown which must be determined
using the condition (\ref{4.45}).\\
The resultant expressions for $b_{11}$ and $b_{22}$ become
\bea      &&b_{11}(q, s)=b_0(q)+b_1(q)s+b_2(q)s^2 \nn\\
         &&b_{22}(q, s)=b_0(q)+b_1(q){1\over s}+b_2(q){1\over{s^2}}
\label{4.47} \eea
 with $q$-dependent coefficients as 
\bea
   b_0&=&{{a^2}\over{4c^2}}-q{a\over{4c}}+{1\over c}(1-{1\over f})\nn\\
   b_1&=&-{{2{\r}q}\over{fc}}\nn\\
   b_2&=&-{{{\s}+{\t}q^2}\over{fc}}
\label{4.48} \eea
Now for $b_{\l\l}(q, s)$ to be the square of a polynomial (in $s$ or $1/s$),  
we must have
\be    b_1^2(q)=4b_0(q)b_2(q)
\label{4.50} \ee 
For this to be held identically the relations among coefficients must be
\bea      &&{\v}^2=1\nn\\
          &&{\s}={\t}={\r}^2/2\nn\\
          &&{\k}=-2
\label{4.51} \eea
which leads to\ft{ We must discard the 
possibility of ${\ct}' =-{\ct}$ as will be seen below.}
\be       {\ct}' ={\ct}   
\label{tt} \ee
\be      {\bet}{\bO}{\bet}+2{\bs}.{\bet}=-2
\label{k} \ee 
The intersection rule (\ref{4.8}),  combined with relation (\ref{tt}),  
leads to the interesting conclusion:\\
A pair of identical supersymmetric branes in any dimension make a  
a marginal configuration, if the two branes
intersect {\it at two equal} or equivalently {\it at SU(2)} angles.\\
The existence of a configuration of two intersecting branes at SU(2) angles
is a result of its unbroken 1/4 supersymmetry first established
in \cite{41}. A similar conclusion was reached in the context of
string theory calculations in \cite{8},  
where it was shown that the interaction potential
between two $D$-branes at SU(2) angles identically vanishes. Our
result establishes their conclusions in a purely field theoretic
context. The advantage of such a 
derivation is its independence of the high-energy 
model of the theory and in particular of the spacetime dimension.\\
We will return to the other result,  
equation (\ref{k}) shortly. It is suitable here to
emphasize on another feature of the integrability conditions which is
only implicit in our calculations,  i.e. their role in fixing
the suitable $H$-basis for using with the constraints (\ref{4.29}).
This relation becomes clear,  
recalling the role of $w_{{\l}{\lp}}$'s (see (\ref{4.20}))
in our equations.
These are $2^{nd}$ degree homogeneous polynomials in
$h_{rs}$ variables, and have six terms generally. 
However, by choosing a suitable $H$-basis,
we can remove two of these terms
simultaneously.  
For example the expression for
$w_{11}$ has the form 
\be      w_{11}=({\ch}_1^{34})^2(h_{33})^2+
2{\ch}_1^{34}({\ch}_1^{14}+{\ch}_1^{32})h_{33}h_{13}+...
\label{w11} \ee 
where the doted terms do not contain any 
powers of ${\ch}_1^{34}$. We can suppress
the first two terms by choosing an 
$H$-basis for which ${\ch}_1^{34}=0$. As a
result the terms of higher degree 
than $s^2$ (i.e. $qs^3$ and $s^4$) will be
removed from $f_{11}(q, s)$,  so 
that finally a quadratic expression (in $s$) 
for ${b}_{11}(q, s)$ is obtained.
If we choose a basis for which
${\ch}_2^{12}=0$, a similar result for ${b}_{22}(q, s)$ emerges.
These two conditions 
(among the others stated below (\ref{4.17})) 
then determine,  as we
have seen,  the matrix $(a_{{\k}{\l}})$ defining the suitable basis.
The important point here is that 
without taking: ${\ch}_1^{34}={\ch}_2^{12}=0$, 
our equations (\ref{4.41}) for $q(X)$ would not be integrable.
The suitable basis,  using (\ref{4.18}) 
and ${\th}' ={\th}$,  is therefore
\bea    H_1&=&1+h_1+{\cbt}h_2 \nn\\
        H_2&=&1+{\cbt}h_1+h_2
\label{4.53} \eea

{\bf The solutions for $q(X)$ and ${\fa}(X)$}\\
Using ${\th}={\th}'$ in our equations causes considerable simplifications.
A straightforward calculation in this case, using
(\ref{4.25}) and (\ref{4.39}), shows that  
\bea              f(q)&=&(1-q^2)^2\nn\\
                 a(q)&=&-{{2q}\over{1-q^2}}\nn\\
                 a_1(q)&=&a_2(q)=-{{q(1+q^2)}\over{2(1-q^2)}}\nn\\
                 b_{11}(q, s)&=&{{({\a}s+q+{\a}sq^2)^2}\over{(1-q^2)^2}}\nn\\
                 b_{22}(q, s)&=&{{({{\a}\over s}+q+{{\a}\over s}q^2)^2}
                 \over{(1-q^2)^2}}\nn\\
                 b_{12}(q, s)&=&{{({\a}s+q+{\a}sq^2)({{\a}\over s}+q
                 +{{\a}\over s}q^2)}\over{(1-q^2)^2}}
\label{4.55} \eea
where ${\a}\;{\;\=\;}\;{\ct}/(1+{\cbt})$, 
and we have used the numerical relations 
(\ref{4.46}) with $\k =-2$. It is clear
from these expressions that the 
consistency condition (\ref{4.42}) is automatically 
satisfied. Note, however, that if we chose
the other possibility from (\ref{4.51}):
$\ct ' =-\ct $, then
the consistency condition would not be  satisfied.
Using these in the equation(s) (\ref{4.41}) leads to four possibilities
for solutions of $({\dot\p}_1 q, {\dot\p}_2 q)$,  
from which only the ${\ve}_1={\ve}_2=-1$ case
is integrable. This corresponds to
\bea    {\pad}q&=&\2q-{\a}s\nn\\
        {\pbd}q&=&\2q-{{\a}\over s}
\label{4.56} \eea
where $s=e^{-1/2(X_1-X_2)}$ as in (\ref{4.40}). 
This system with the boundary
condition: $q(0, 0)={\ct}$ has a unique solution,
\bea               q&=&{\a}[2cosh (X_1-X_2)/2-\b e^{(X_1+X_2)/2}] \nn \\
                    &=&{\a}(H_1+H_2-{\b})/\sqrt{H_1H_2}
\label{4.57} \eea 
where ${\b}\;{\;\=\;}\;{\sbt}$.\\
The solutions for $\fa (X)$'s,  using this solution for $q(X)$,  
are already expressed as in
(\ref{4.31}) up to the unknown parameters $\ea$.
To determine $\ea$'s (as well as $\xla$'s in principle),  we have to set the
relations between parameters,  required by the consistency of the equations
governing $(\fa (X), q(X))$ and the ansatz (\ref{4.31}) for $\fa (X, q)$.
Four of these relations have been given 
previously by (\ref{4.32}) and (\ref{k}).
For obtaining a fifth one,  we use the embedding
equation of the dilaton,
\be      ({\pld}{\plpd} +{\dllp}{\pld}){\vf}+{\a}_0{\fl}/f=0
\label{4.59} \ee
where have used the constraints (\ref{4.29}) to eliminate the exponentials.
On the other hand, the dilaton (i.e. $\a =5$) 
component of (\ref{4.31}) is written as
\be        {\vf}=-{\a}_0(X_1+X_2)+\eta^5\ln \sqrt{f(q)/\g}
\label{4.60}   \ee
where we have put ${\xi}_1^5={\xi}_2^5={\a}_0$ using (\ref{1.20}).
Inserting this ansatz in (\ref{4.59}),  
a new set of equations involving only $q(X)$ 
results. For these to become identities,  
equations (\ref{4.56}) require that:
$\eta^5={\a}_0$. Putting this value in 
the equations (\ref{4.32}) and (\ref{k})
and solving for $\ea$'s, then gives 
\bea   
{\ea}&=&(-{{{\dt}_0}\over{2{\Db}}}, 
-{{{\dt}_0}\over{2{\Db}}}, 
-{{{\dt}_0}\over{2{\Db}}}, {{d_0}\over{2{\Db}}}, {\a}_0) \nn \\
\label{4.62}  \eea
This completes our solution (\ref{4.31}) for $\fa (X)$.\\

{\bf The solutions for $\cA (X)$}\\
For purely electric-type branes only the  `world-volume'  components of the
form-potentials survive (see (\ref{3.1})),
\be                   {\cA}=A_{({\m})rs}({\dx}){\w}dy^r{\w}dy^s
\label{4.63} \ee 
Applying the formula (\ref{3.18}) to this case, we obtain

\be      {\pl}A_{({\m})rs}=
{\e}_{({\m})}{\ch}_{\l}^{r' s' }h_{rr' }h_{ss' } 
e^{2({\d}B_0-\hat{G}-{\a}_0{\vf})}
\label{4.64} \ee 
With the help of (\ref{4.29}), and in the form-notation, eq. (\ref{4.64}) is
written as
\be           dA_{({\m})rs}={\e}_{({\m})}{\o}_{rs}
\label{dA} \ee
where ${\o}$'s are a set of 1-forms (on the $H$-surface) defined by
\be  {\o}_{rs}\;{\;\=\;}\;{{f_{{\l}(rs)}}\over f}{\Hl}^{-2}d{\Hl}
\label{4.68} \ee 
where $\flrs$'s are functions of $(q,s)$ with the non-vanishing members:
\bea
&&f_{1(12)}=1+2{\a}sq\;\;\;\;\;\;\;\;\;\;\;\;\;\;\;\;\;\;\;\;\;\;\;
\;\;\;\;\;\;\; f_{2(12)}=2{\a}{q\over {s^3}}+{{q^2}\over{s^2}}\nn\\
&&f_{1(34)}=2{\a}s^3q+s^2q^2\;\;\;\;\;\;\;\;\;\;\;\;\;\;\;\;\;\;\;\;
\;\;\;\; f_{2(34)}=1+2{\a}{q\over s}\nn\\
&&f_{1(32)}=f_{1(14)}{=}{\a}s^2+sq+{\a}s^2q^2\;\;\;\;\;\; 
f_{2(32)}{=}f_{2(14)}={{\a}\over{s^2}}+{q\over s}+{\a}{{q^2}\over{s^2}}
\label{4.67} \eea 
For the equations (\ref{dA}) (for $A_{(\m )rs}$'s) 
to be integrable,  the forms
${\o}_{rs}$ must be exact, which can be checked by expressing them totally
in $(q,s)$ variables. Therefore $\o$'s can be integrated to yield,
\be      e_{rs}=-\int {\o}_{rs}
\label{4.70} \ee
The non-vanishing $e_{rs}$'s are;
\bea      e_{12}&{\;\=\;}&{{1+{1\over{s^2}}
+{q\over{{\a}s}}}\over{{\b}(1-q^2)}}\nn\\
          e_{34}&{\;\=\;}&{{1+s^2+{{sq}\over{\a}}}\over{{\b}(1-q^2)}} \nn\\
          e_{32}&{\;\=\;}&e_{14}{\;\=\;}{{(s+{1\over s})q
          +{{q^2}\over{\a}}}\over{{\b}(1-q^2)}}
\label{4.71} \eea 
Finally the solutions for the components of $\cA$ are written as
\be      A_{{\m}(rs)}=-{\e}_{({\m})}e_{rs}
\label{4.72} \ee

{\bf Summary of the solutions}\\
In the following we present the final 
forms of the solutions in terms of the
variables $({\Hl}, q)$,  without 
inserting the $H$-dependences of $q$ in them.
This choice makes them more transparent. First, 
the expressions for $ds^2$ 
and $e^{{\vf}}$ are found from (\ref{4.31}),  using (\ref{1.20})
and (\ref{4.62}),  as
\bea  ds^2&=&\left (\sqrt{f(q)/\g}H_1H_2\right )^{-{\dt}_0
/{\Db}}[{\dx}{\dxp}+H_2(dy_1^2+dy_2^2)+H_1(dy_3^2+dy_4^2)\nn\\
&&+2q\sqrt{H_1H_2}(dy_1dy_3+dy_2dy_4)]
+\left (\sqrt{f(q)/\g}H_1H_2\right )^{d_0/{\Db}}{\dz}{\dz}
\label{4.73} \eea 
\be    e^{\vf}=\left (\sqrt{f(q)/\g}H_1H_2\right )^{{\a}_0}
\label{4.74} \ee 
Second,  the expression for ${\cA}$ is 
found,  from (\ref{4.71}) and (\ref{4.72}),  as
\be       
{\cA}=-{1\over\sqrt{f(q)/\g}}({\dx}){\w}[{1\over{H_1}}dy^1{\w}dy^2
+{1\over{H_2}}dy^3{\w}dy^4+{q\over{\sqrt{H_1H_2}}}(dy^1{\w}dy^4
+dy^3{\w}dy^2)]
\label{4.75} \ee 
In these formulas,  $H$-dependence 
of $q$ is given by (\ref{4.57}) and $f(q)$ is defined in
(\ref{4.55}). The harmonic functions $(H_1(z), H_2(z))$ are related to
the charge densities $({\r}_1(z), {\r}_2(z))$ 
via (\ref{3.40}) and (\ref{4.53}) which
altogether may be written as
\bea    H_1(z)&=&1+{\izp}G_{{\dt}+2}(z, z' )
[{\r}_1(z' )+{\cbt}{\r}_2(z' )]\nn\\
        H_2(z)&=&1+{\izp}G_{{\dt}+2}(z, z' )[{\cbt}{\r}_1(z' )+{\r}_2(z' )]
\label{4.76} \eea
By expressing $(H_1, H_2)$ in terms of $(h_1, h_2)$, 
it is easy to check that for Type IIA $(2,2)$-branes 
with $D=10$ and $d_0=3$, this solution
exactly matches that of \cite{22} expressed as (\ref{3.32}) previously. 
In particular the function $E$ in that solution is nothing but:
$E=\sqrt{f(q)/\g}H_1H_2 $.
The general case with arbitrary $D$ and $p$ had not been reported earlier.\\
\section{No-force conditions}
\setcounter{equation}{0}
No-force conditions are a set of 
constraints arising
naturally when we consider the marginally 
stable (static or stationary) configurations of
\pbs. We have seen how a class of field 
theory solutions for these systems
are constructed based on 
a very special set of
constraints corresponding to the 
extremality and no-force conditions.
The interpretation of the former was given in section 3. 
In this section we give a precise interpretation of the latter,  
which justifies
its name.
We start from a formulation of the \nfc similar to that of \cite{10}.
In principle to find such conditions,  
one has to single out every constituent
$({\dl}-1)$-brane from rest of the branes,  and demand its  `equilibrium' 
conditions under interactions with the others. Equivalently we may look for
the equilibrium conditions 
of a $({\dl}-1)$-brane  `probe' ,  with nearly zero
mass and charge,  situated parallel 
to the similar distribution
within the system. Since  
the spacetime geometry along the world-volume
of the brane in equilibrium is homogeneous,  we 
can decompose the metric as
\be      ds^2=h_{{\m}{\n}}(y)dx^{\m}dx^{\n}+k_{mn}(y)dy^mdy^n
\label{5.1} \ee
where $(x^{\m})$ and $(y^m)$ stand for the coordinates 
parallel and transverse to
the world-volume of the $({\dl}-1)$-brane respectively. 
The action governing the
dynamics of the brane is the DBI action.  
Since our branes have no boundaries, 
we can truncate DBI actions such that all the internal gauge 
fields of the branes vanish \cite{60}. Therefore the
static gauge action of a $(\dl -1)$-brane with $X^{\m}={{\xi}}^{\m}$ ,  
$Y^m=Y^m({\xi})$ is written as
\be
S_{\l}[Y]=-T_{\l}\int d^{{\dl}}{\xi}\{e^{-{\al}{\vf}(Y)}
[-det\left (h_{{\m}{\n}}(Y)+
k_{mn}(Y){\p}_{\m}Y^m{\p}_{\n}Y^n\right )]^{1/2}
+{\kl}^{-1}{\e}_{\l}^{({\m})}A_{\l (\m )}(Y)\}   
\label{5.2} \ee
Here $T_{\l}$ is the brane tension,  
${\e}_{\l}^{({\m})}$ the ${\dl}$-dimensional alternating
symbol, and $A_{\l (\m )}$ represents the pull-back of the $\dl$-form
potential on the world-volume of the $(\dl -1)$-brane. 
In (\ref{5.2}) we have introduced an unusual factor
${\kl}$, which must be the same as the mass to charge ratio (\ref{1.23})
of the brane (see below). 
For ${\kl}={\pm}1$ we recover the usual
action of super \pbs. Expanding this action in 
powers of the velocities: $({\p}_{\m}Y^m)$
and keeping only the leading order terms,  
we obtain the static potential between a
$({\dl}-1)$-brane probe and the complete brane system as
\be    V_{\l}(Y)=T_{\l}\left 
(e^{-{\al}{\vf}(Y)}\sqrt{h_{\l}(Y)}
+{\kl}^{-1}{\el}^{({\m})}A_{{\l}({\m})}(Y)\right )
\label{5.3} \ee
where $h_{\l}\;{\;\=\;}\;|det(h_{{\m}{\n}})|$. In the equilibrium
state,  we must have $V_{\l}(Y)=$ constant. Eliminating this constant 
by a suitable gauge transformation on ${\cA}_{\l}$, the no-force condition
for a $(\dl -1)$-brane is written as
$V_{\l}=0$ , alternatively
\be    e^{-{\al}{\vf}}h_{\l}^{1/2}
=-{\kl}^{-1}{\el}^{({\m})}A_{{\l}({\m})}
\label{5.4} \ee
Note that the above equation gives 
explicitly the component of a $\dl$-form
potential parallel to $({\dl}-1)$-brane,  
as a function of the dilaton and the
(determinant of the) corresponding metric 
components. This can be compared to
an analogous but different result,  
which was given in the form of equation
(\ref{3.18}) previously.
As the the number of the equations in (\ref{5.4}) 
is equal to the number of constraints (\ref{3.78})
, $N$, we may tend to guess that these two sets of equations are
equivalent. This is in fact true  for all the
configurations considered earlier, as will be seen below.\\

{\bf The case of $N$ orthogonal branes}\\
In this case by (\ref{2.2}),  for all the 
branes within the ${\l}^{th}$ distribution, 
we have
\bea     A_{{\l}({\m})}&=&{\e}_{{\l}({\m})}e^{\Xl} \nn \\
         e^{-{\al}{\vf}}h_{\l}^{1/2}
         &=&exp(\sum_i {\d}_{{\l}i}B_i-{\al}{\vf})
\label{5.5} \eea
So noting $e^{\Xl}={\kl}{\Hl}^{-1}$,  equation (\ref{5.4}) gives
\be          \sum_i {\d}_{{\l}i}B_i-{\al}{\vf}=-\ln{\Hl}
\label{5.6} \ee
Note that cancellation of 
the factor ${\kl}$ in this equation, occurs only if 
one introduces the factor
${\kl}^{-1}$ in the action (\ref{5.2}). This 
confirms the result of section 1 that 
$\kl$ is the mass to charge ratio of
a $(\dl -1)$-brane.
We can put (\ref{5.6}) into a more familiar form,  using the
extremality constraint (equation (\ref{3.72})) 
\be         G({\bff})\;{\;\=\;}\;\2\left (\sum_iB_i+{\dt}C\right )=0
\label{5.7} \ee
Combining the last two equations gives
\be         {\Fl}({\bff})\;{\;\=\;}\;\2\sum_i{\ve}_{{\l}i}B_i
-1/2{\dt}C-{\al}{\vf}=-\ln{\Hl}
\label{5.8} \ee
This is exactly the second constraint in (\ref{1.10}) 
or its generalization to (\ref{3.78}).
Clearly this derivation does not rely 
on supersymmetry properties of the theory, as it does not
rely on choosing $\kl =\pm 1$. In this sense,  
the marginality of a solution is not
necessarily a result of its supersymmetries.\\

{\bf The case of $(p, p)$-branes at angles}\\
For comparing with section 4,  we consider only the case of super p-branes.
In this case,  using the ansatz (\ref{4.3}) for $\hij$ and 
the solution (\ref{4.75}) for
${\cA}$,  we obtain
\bea  &&A_{\l}={\Hl}^{-1}{[f(q)/\g]}^{-1/2}\nn\\
      &&e^{-{\a}_0{\vf}}{h}_{\l}^{1/2}=e^{2B_{\l}+{\d}B_0-{\a}_0{\vf}}
\label{5.9} \eea
where we have introduced $A_{\l}\;{\;\=\;}\;{\el}^{({\m})}A_{({\m})}$.  
So (\ref{5.4}) takes the form
\be           2B_{\l}+{\d}B_0-{\a}_0{\vf}=-\ln{\Hl}-\ln\sqrt{f(q)/\g}
\label{5.11} \ee
Again the extremality constraint (first equation in (\ref{4.29})),  combined
with this equation,  gives
\be           {\Fl}({\bff})+1/2\ln{\sqrt{f(q)}}=-\ln{{\Hl}}+1/2\ln{\g}
\label{5.*} \ee
This represents the second constraint in (\ref{4.29}),  as was expected.\\

{\bf Long range potentials}\\
The (short and long range) potentials between D-branes in the
spacetime of dimensionality $D=10$,  
due to exchange of (massive and massless)
closed superstring states,  have 
been calculated in many places \cite{50,51,36}. However
there has not been a general prescription for 
such calculations in spacetimes
of arbitrary dimension,  as a quantum theory of strings in other dimensions
does not make sense. Nevertheless by studying the long 
distance behaviour of the
 `effective'  potential (\ref{5.3}),  
 we can specify the contributions from
the various (massless) bosonic fields in 
our (low energy) model to the total long-range
potential between a brane probe and a system of distributed branes. This
will give physical interpretations to the 
various consistency conditions,  arising
in the discussion of marginal 
solutions (e.g. eqns.(\ref{1.21})). To this end, we
first note that the various dynamical 
variables in these solutions have power
series expansions in terms of the 
harmonic functions ${\D}{\Hl}{\;\=\;}{\Hl}-1$, which all
tend to zero as 
$z{\ra}\infty$\ft{In general $\D\Hl$'s are written as linear
combinations of $\hl$'s defined in (\ref{3.40}). 
But for the orthogonal case considered 
here,  they actually coincide.}.
So in particular (noting the boundary conditions)
we have the expansions
\bea    e^{\vf}&=&1+{\D}{\vf}+...\nn\\
        A_{\l}&=&{\kl}(1-{\D}{\Hl}+...)\nn\\
        h_{\l}&=&1+{\D}h_{\l}+...
\label{5.12} \eea
where the dots stand for higher than 
first order terms in ${\D}{\Hl}$'s. Using these
in (\ref{5.3}) we find the expansion for $V_{\l}(z)$ (to first order) as
\be       V_{\l}=T_{\l}(\2{\D}h_{\l}-{\al}{\D}{\vf}+{\D}{\Hl})+...
\label{5.13} \ee
We identify the three leading order terms of this 
expansion as the  gravitational, 
dilatonic,  and $({\dl}+1)$-form contributions to the potential energy
of a distant $({\dl}-1)$-brane,  which is parallel to 
the corresponding distribution
within the system,  that is
\be
V_{G{\l}}=\2T_{\l}{\D}h_{\l}\;\; , 
\;\;V_{D{\l}}=-T_{\l}{\al}{\D}{\vf}\;\; , \;\;V_{F{\l}}=T_{\l}{\D}{\Hl}
\label{5.14} \ee
Such an identification is possible,  as each of these terms involves
$only$ the variations of the 
corresponding field variables and not mixings among
themselves. As an example we 
consider the case of the orthogonal system $d_1\cap d_2={\d}$
and a $(d_1-1)$-brane probe. Then 
from the solutions (\ref{1.28}) we find
\bea   {\D}h_1&=&-{\k}_1^2{{d_1{\dt}_1}\over{\Db}}{\D}H_1
+{\k}_2^2({{d_1d_2}\over{\Db}}-{\d}){\D}H_2\nn\\
       {\D}{\vf}&=&{\k}_1^2{\a}_1{\D}H_1+{\k}_2^2{\a}_2{\D}H_2
\label{5.15} \eea
Specializing to the case of 
localized distributions along the transverse space 
at the points $z_{\l}^a$ (${\l}=1, 2$), we have
\be     {\D}H_{\l}(z)={{c_{\l}T_{\l}}\over{|z-z_{\l}|^{\dt}}}
\label{5.16} \ee
where $c_{\l}$'s are constants proportional 
to the densities of the corresponding
longitudinal distributions. 
The potentials (\ref{5.14}) for e.g. a $(d_1-1)$-brane
probe at point $z^a$ thus become
\bea   
V_{G1}(z)&=&-c_1{\k}_1^2{{d_1{\dt}_1}\over{\Db}}{{T_1^2}\over{|z-z_1|^{\dt}}}
+c_2{\k}_2^2({{d_1d_2}\over{2{\Db}}}
-{{{\d}}\over 2}){{T_1T_2}\over{|z-z_2|^{\dt}}}\nn\\
V_{D1}(z)&=&-c_1{\k}_1^2{{{\a}_1^2T_1^2}\over{|z-z_1|^{\dt}}}
-c_2{\k}_2^2{{{\a}_1{\a}_2T_1T_2}\over{|z-z_2|^{\dt}}}\nn\\
V_{F1}(z)&=&c_1{{T_1^2}\over{|z-z_1|^{\dt}}}
\label{5.17} \eea
where the first (second) term in each expression represents the interaction
potential with the $(d_1-1)$-branes ($(d_2-1)$-branes) distribution. 
As these 
formulas indicate,  the gravitational 
and dilatonic forces between two same-type 
(parallel) branes are always attractive,  while their  form-field 
force is restrictly repulsive. In contrast,  both the gravitational 
and dilatonic
forces between two (orthogonal) 
different-type branes may be either attractive or 
repulsive depending on the values of 
the various dimensions and couplings,  but 
there are no form-field forces 
between themselves,  as is expected. When the 
$(d_1-1, d_2-1)$-branes system is 
in a marginal (BPS) state,  the internal branes as 
well as the brane probe do not 
feel any total force. So at an arbitrary point
$(z^a)$ we must have $V_1(z)=V_{G1}(z)+V_{D1}(z)+V_{F1}(z)\;{\;\=\;}\; 0$,  
which according
to (\ref{5.17}) requires
\bea    -{\k}_1^2{{d_1{\dt}_1}\over{2{\Db}}}-{\k}_1^2{\a}_1^2+1&=&0\nn\\
        ({{d_1d_2}\over{2{\Db}}}-{{{\d}}\over 2})-{\a}_1{\a}_2&=&0
\label{5.18} \eea
Interchanging the roles of $(d_1-1)$ and $(d_2-1)$-branes in (\ref{5.18}), 
we obtain
the three consistency conditions (\ref{1.21}).
This verifies explicitly our assertion in section 1 that the consistency 
conditions are nothings but the requirements of the no-force 
conditions between
different pairs of branes.\\

\section{Masses and charges}
\setcounter{equation}{0}
The important physical 
parameters of a brane configuration (appearing e.g. in the black hole 
applications) are its mass and various charges.
The total mass is the sum of the (positive) rest masses
of the constituent branes,  and the (negative) energy of the
binding forces, both contained in the ADM mass of the brane system.
For a marginal configuration, the $total$ interaction energy vanishes and
one expects that the ADM
mass to be the sum of the constituents masses. The general ADM mass formula
is derived by linearizing the Einstein equation in 
the (flat) background of the
asymptotic metric \cite{28}. For a 
system of distributed branes with a metric of the
form (\ref{3.1}),  the general formula 
(in the units with $16{\pi} G_D=1$) reduces to
\be          {\cM}\;{\;\=\;}\;{M\over{V_{d-1}}}=-{\iz}{\p}^2\Psi (z)
\label{6.5} \ee
where $V_{d-1}$ is the volume of 
the distribution subspace,  and $\Psi (z)$
is
\be           
\Psi (z)\;{\;\=\;}\;{\gijp}{\hij}(z)+h_{00}(z)+({\dt}+1)e^{2C(z)}+const.
\label{6.6} \ee
In this formula,  
$\gijp\;\=\;\hijp (z) |_{z\ra\infty}$. Using the Stokes theorem, 
the expression for ADM mass becomes
\be  {\cM}=-{\int}_{S^{{\dt}+1}}d^{{\dt}+1}{{{\O}}}
r^{{\dt}+1}{\p}_r\Psi (r, {\th}^{\dot{a}})
=-{\O}_{{\dt}+1}{(r^{{\dt}+1}{\p}_r\Psi )}_{r{\ra}\infty}
\label{6.7} \ee
where $(r, {\th}^{\dot{a}})$ 
denotes a set of polar coordinates on the transverse 
space,  and ${\O}_{{\dt}+1}$ stands for the unit $({\dt}+1)$-sphere's area.
From this formula a finite value for ${\cM}$ is obtained,  
only if $\Psi$ at infinity behaves
as ${1\over{r^{\dt}}}$ (i.e. as an harmonic function). 
In fact by expressing the metric
in terms of the harmonic functions ${\hl}$ 
as in (\ref{3.40}),  and expanding
in powers of ${\hl}$'s we obtain
\be      \Psi ({\hl})=\sum_{\l}{\m}_{\l}{\hl}+{\cal O}(h_{\l}^2)
\label{6.8} \ee
where ${\m}_{\l}$'s are constants depending 
on the specifications of the system
under consideration. Using this in (\ref{6.7}),  
and reversing the route of Stokes
theorem from (\ref{6.7}) to (\ref{6.5}),  
yields 
\be       {\cM}=-\sum_{\l}{\m}_{\l}{\iz}{\p}^2{\hl}
=\sum_{\l}{\m}_{\l}{\iz}{\rl} (z)
\label{6.9} \ee
where we have used the Poisson equations 
${\p}^2{\hl}=-{\rl}$ within the distribution
region of the transverse space. 
Note that, despite using only the
${\cal O}(\hl )$ terms of (\ref{6.8}),  
this is an exact expression for $\cM$, since 
higher order terms of $\Psi$ do not 
contribute to the surface integral in (\ref{6.7}).\\

The charges are determined by types of the existing branes.
In this paper we deal with cases where only $electric$ charges present.
The $total$ electric charge,  corresponding 
to a $(d_r+1)$-form field strength ${\cF}^r$, 
is defined via a $d_r$-form conserved current ${\cJ}^r$ as
\be
Q^r=\int_{V^{{\dt}_r+2}}*{{\cJ}}^r
=\int_{S^{{\dt}_r+1}}e^{2{\a}_r{\vf}}*{\cF}^r
\label{6.1} \ee
where $V^{{\dt}_r+2}$ is any 
$({\dt}_r+2)$-dimensional hyperplane intersecting all
the existing $(d_r-1)$-branes at 
points and $S^{{\dt}_r+1}$ is a $({\dt}_r+1)$-sphere
surrounding these points. The * in (\ref{6.1}) 
denotes the Hodge dual in the
curved background. By taking the radius of $S^{{\dt}_r+1}$ to infinity, 
the fields in (\ref{6.1}) are replaced by 
their asymptotic values and it reduces to
\be
Q^r=\int_{S^{{\dt}_r+1}}*{{\cF}}_\infty^r
=\int_{V^{{\dt}_r+2}}d*{{\cF}}^r_\infty 
\label{6.2} \ee
where now * denotes the Hodge dual with 
respect to the flat background,  and we
have set as in the previous sections ${\vf}_{\infty}=0$. 
Using the ${\cA}^r$'s equation of motion
at infinity (see section 3)
\be     
d*{{\cF}}^r_\infty=*{\cJ}^r_\infty
=\sum_{{\dl}=d_r}{\kl}{\rl}(z)*{\el}
\label{6.3} \ee
eq. (\ref{6.2}) becomes
\be       {\cQ}^r\;{\;\=\;}\;{{Q^r}\over{V_{d-d_r}}}
=\sum_{{\dl}=d_r}{\kl}{\iz}{\rl}(z)
\label{6.4} \ee
where $V_{d-d_r}$ denotes the volume of 
the subspace transverse to $(d_r-1)$-brane's
world-volume and parallel to the 
distribution subspace. This shows that the charges of the
same-type branes are additive 
even if they are not parallel,  and ${\kl}{\rl}(z)$
measures the charge density of 
the ${\l}^{th}$ distribution per its unit
transverse volume. We examine in 
the following the general formulas (\ref{6.4})
and (\ref{6.9}) for the orthogonal 
solution of section 2. Similar results hold for the non-orthogonal
solution of section 4.\\

{\bf The case of $N$ orthogonal branes}\\
In this case from the ansatz (\ref{2.2}), 
and using (\ref{1.10}) and ({1.12}), we have
\bea  e^{2{\al}{\vf}}*{\cF}_{\l}&=&
      e^{-2{\Fl}({\bff})}\tilde{*}de^{\Xl}{\w}
      \left ({\W}_{i\notin{\dl}}dx^i\right )\nn\\
    &=&{\kl}\tilde{*}d{\Hl}{\w}\left ({\W}_{i\notin{\dl}}dx^i\right )
\label{6.10} \eea
where the $\tilde{*}$ denotes the 
Hodge dual in a $({\dt}+2)$-dimensional Euclidean
space (an irrelevant overall sign concerning 
dualization have been omitted).
 Using this in (\ref{6.1}), 
we obtain in agreement with (\ref{6.4}) that
\be   {\cQ}_{\l}={\kl}\int_{S^{{\dt}+1}}\tilde{*}d{\Hl}
={\kl}\int_{V^{{\dt}+2}}d^{{\dt}+2}z{\rl}(z)
\label{6.11} \ee
where the Stokes theorem with and Poisson equation have been used.\\
For the mass formula,  using (\ref{6.6}) 
with the solution (\ref{2.12}),  we obtain
\be          \Psi =\sum_{\l}{\kl}^2{\D}{\Hl}+{\cal O}({\D}H^2)
\label{6.12} \ee
which shows that ${\m}_{\l}={\kl}^2$ in this case. So by (\ref{6.9}) we have
\be        {\cM}=\sum_{\l}{\kl}^2{\iz}{\rl}(z)=\sum_{\l}|{\kl}{\cQ}_{\l}|
\label{6.13} \ee
That is the total mass $\cM$ equals the sum of the constituent masses
${\cM}_{\l}=|{\kl}{\cQ}_{\l}|$, which is an indication of marginality.\\

{\Large{\bf Conclusion}}\\
Starting from a reduced Lagrangian 
reformulation of the problem of orthogonal
brane solutions,  we arrived at a 
set of (linear) constraints,  which was shown to consistently
solve the corresponding (non-linear) field equations. 
The requirement of consistency between
these two sets of equations,  
led to a set of algebraic constraints containing
all the physical information 
characterizing the marginal orthogonal solutions.
These include the mass to charge ratios of the constituent branes and their
suitable intersection rules. Although in the 
realistic supergravities this lead
to the BPS saturated solutions with 
extremal (super) \pbs as the building blocks, 
extensions to the
black solutions with non-extremal (black) \pbs 
are also possible using suitable
deformation functions \cite{16}. 
By introducing a general formulation for handling
arbitrary geometries of the 
intersecting branes with uniform  `longitudinal' 
distributions,  a very general 
expression for the associated form-potentials in terms of the 
metric and dilaton field
was derived. It was shown that 
the equations of motion of the reduced theory, 
can be translated to the  `forced-geodesic'  
equations describing a surface
in the fields  `configuration space' . 
The conditions for the integrability of these
equations are found to coincide with the 
constraints obtained earlier \cite{37}.
Essentially this type of formulation is 
not restricted to the case of the marginal
solutions,  as far as the number of 
the independent harmonic functions is not
restricted to that of the density functions. 
The distributions densities may be
so correlated to result in the dependent harmonic 
functions. Consequently the
constraints of the marginal solutions are not valid for the
non-marginal solutions. As a 
result the suitable intersection rules will be different
from those of the marginal solutions. 
We hope that the formulation of this paper
(with suitable changes) to 
be applicable for classifying these non-marginal solutions
as well (see however \cite{52}). 
Finally we showed that how applying the ideas 
of the $H$-surface and null geodesic surface lead to 
the solutions for a system with 
two similar branes at $SU(2)$ angles.\\

{\Large{\bf Appendix:}}\\
\renewcommand{\theequation}{A.\arabic{equation}}
\setcounter{equation}{0}
{\large{\bf A. First order RL's for gravity}}\\
The standard Einstein-Hilbert action is written
in terms of the $2^{nd}$ order Lagrangian: 
\be    {\cL}_{EH}=\sqrt{-g}R(g,\p g,{\p}^2g)
\label{A.1} \ee
However, that the Einstein equation itself is of second
order, shows that  $\cL_{EH}$  must be `equivalent'  
to a $1^{st}$ order Lagrangian.  
To see this explicitly, we apply the formula for Ricci tensor:
\be     
R_{MN}={1\over\sqrt{-g}}{\p}_P(\sqrt{-g}{\G}_{\;MN}^P)
-{\G}_{\;QM}^P{\G}_{\;PM}^Q-{\p}_M{\p}_N(\ln\sqrt{-g})   
\label{A.2} \ee
where the Christoffel symbols, ${{\G}}_{\;MN}^P$ 's are 
$1^{st}$ order quantities. Using this formula,  the expression for
$\sqrt{-g}R$ becomes
\bea   
\sqrt{-g}R  =  &&g^{MN}{\p}_P(\sqrt{-g}{\G}_{\;MN}^P)
-\sqrt{-g}{\G}_{\;QM}^P{\G}_{\;PM}^Q
-\sqrt{-g}{\p}^M{\p}_M(\ln\sqrt{-g}) \nn\\              
                   = &&-\sqrt{-g}{\p}_Pg^{MN}{\G}_{\;MN}^P
                   -\sqrt{-g}{\G}^{PQM}{\G}_{QPM}
                   +{\p}_M(\ln\sqrt{-g}){\p}_N(\sqrt{-g}g^{MN}) \nn\\
                     &&+{\p}_P(\sqrt{-g}g^{MN}{\G}_{\;MN}^P
                     -{\p}_N(\sqrt{-g}g^{MN}{\p}_M\ln\sqrt{-g})
\label{A.3} \eea
where the $2^{nd}$ order terms in the first line,  have been appeared
as total derivatives in the third line. 
After further simplification using the
formulas
\bea           {\p}_Pg^{MN}{\G}_{\;MN}^P&=&-2{\G}^{PQR}{\G}_{QPR}\nn\\
                   g^{MN}{\G}_{\;MN}^P&=&-{\p}_Qg^{PQ}-{\p}^P(\ln\sqrt{-g})
\label{A.4} \eea
equation (\ref{A.3}) reduces to
\be          \sqrt{-g}R={\cL}_G(g, {\p}g)-{\p}_M{\L}^M(g, {\p}g)
\label{A.5} \ee
where we have defied
\bea     {\cL}_G&\=&\sqrt{-g}{\G}^{PMN}{\G}_{(MN)P}
+{\p}_M(\ln\sqrt{-g}){\p}_N(\sqrt{-g}g^{MN})\\
\label{A.6} 
          {\L}^M&\=&\sqrt{-g}({\p}_Ng^{MN}+2g^{MN}{\p}_N\ln\sqrt{-g})
\label{A.7}  \eea
Formulas (\ref{A.5}) to (\ref{A.7}) can serve  
as simplifying explicit formulas for 
practical calculations of the Ricci scalar. 
However, since the surface term in (\ref{A.5})
does not  contribute to the equations of motion, we can identify 
the $1^{st}$ order part, $\LG$, as the gravitational field's
Lagrangian.\ft{It must be noted however that ${\cL}_G$ despite ${\cL}_{EH}$
is not transforming as a scalar density 
under coordinate transformations.}\\.
 
{\bf Application to semi-homogeneous spacetimes }\\
We have constructed examples of such 
spacetimes using the brane distributions
throughout sections 1 to 4 of this paper. 
The generic form of the metric tensor
for such spacetimes can be written as 
(see (\ref{3.1}) and the related descriptions)
\be       g_{MN}=\pmatrix{{\hij}(z)&0\cr 0&{\d}_{ab}e^{2C(z)}}
\label{A.8} \ee
From this metric we have
\be       \ln\sqrt{-g}=1/2\ln h+({\dt}+2)C
\label{A.9} \ee
where $h{\;\=\;}|det({\hij})|$. The non-vanishing Christoffel symbols are
\bea        {\G}_{ija}&=&-{\G}_{aij}=1/2{\p}_a{\hij}\nn\\
            {\G}_{abc}&=&e^{2C}({\d}_{ab}{\p}_cC
            -{\d}_{bc}{\p}_aC+{\d}_{ca}{\p}_bC)
\label{A.10} \eea
The expression for ${\LG}$ in (\ref{A.6}) thus becomes
\be    {\cL}_G=h^{1/2}e^{{\dt}C}[1/4{\p}_a{\hij}{\p}_a{\hijp}
+1/4({\p}_a\ln h)^2+{\dtt}({\p}_a\ln h){\p}_aC+{\dt}{\dtt}({\p}C)^2]  
\label{A.11} \ee
After simple manipulations,  
using the matrix notation,  this formula is rewritten as
\be    {\cL}_G=e^{2G}[-1/4{\tr}^2
+1/4{\trb}+{\dtt}{\p}C.{\tr}+{\dt}{\dtt}({\p}C)^2]
\label{A.12} \ee
Here we have dropped the (contracted) 
transverse space indices,  taken traces over
$(i, j)$  and defined $G$ as
\be     2G\;{\;\=\;}\;1/2\ln h+{\dt}C
\label{A.13} \ee

{\bf A simplifying trick for calculating ${\cL}_G$}\\
Finding a closed form for ${\cL}_G$ 
in the specific problems (given the ansatz 
for ${\hij}$) by (\ref{A.12}) needs to a 
closed form of ${\hijp}$,  which 
in many situations can not be found easily. 
Fortunately a shortcut exists
by means of which ${\cL}_G$ can be computed 
without really inverting ${\hij}$.
All that is needed,  is to calculate 
the determinant: $h=det({\hij})$,  and
take its $1^{st}$ and $2^{nd}$ 
variations. This originates from the formula
\be        {\hijp}={{{\p}(\ln h)}\over{{\p}{\hij}}}
\label{A.14} \ee
using which the two traces in (\ref{A.12}),  take simple expressions as
\bea        
&&{\tr}={{{\p}(\ln h)}\over{{\p}{\hij}}}{\p}{\hij}={\p}(\ln h)\nn\\
&-&{\tr}^2={{{\p}^2(\ln h)}\over{{\p}{\hij}{\p}h_{kl}}}\p h_{ij} .\p h_{kl}
={({\p}^2\ln h)}_{{\p}^2{\hij}=0}
\label{A.15} \eea
That is the two traces are calculated by expressing $\ln h$ as
a function of $({\hij})$,  taking its $1^{st}$ 
and $2^{nd}$ order variations,  and neglecting
the $2^{nd}$ order variation of ${\hij}$ 
\ft{When calculating 
the second variation,  we are allowed to replace 
${\hij}$'s with another set of 
variables in terms of which ${\hij}$'s are linear.}.
Finally these two expressions are 
combined into the equation
\be     {\trb}-{\tr}^2
={\left ({{{\p}^2h}\over h}\right )}_{{\p}^2{\hij}=0}
\label{A.17} \ee
This is proved to be a very useful 
formula for the sake of practical calculations.\\

{\bf Application to the system $d_1\cap d_2=\d$ at angles}\\
For this system ${\hij}$ is defined by (\ref{4.1}). 
Computing the determinant of
this matrix we obtain
\be   h=f(q)e^{2({\d}B_0+{\d}_1B_1+{\d}_2B_2)}
\label{A.18} \ee
where
\be     f(q){\;\=\;}|det(1-q^2{\bga}{\bga}^T)|
\label{A.19} \ee
Defining the four independent variables $q^A$ linear in ${\hij}$'s as
\be     q^A{\;\=\;}(e^{2B_0}, e^{B_1}, e^{2B_2}, qe^{B_1+B_2})
\label{A.20} \ee
and using the above tricks,  with ${\p}^2q^A=0$ in (\ref{A.17}),  
we obtain
\bea      
&&1/2{\tr}={\d}{\p}B_0+{\d}_1{\p}B_1
+{\d}_2{\p}B_2+{{f' }\over{2f}}{\p}q\nn\\
            &&1/4[{\trb}-{\tr}^2]=\nn\\
            &&{\d}({\d}-1)({\p}B_0)^2
            +[{\d}_1({\d}_1-1)+{{qf' }\over{4f}}]({\p}B_1)^2
            +[{\d}_2({\d}_2-1)+{{qf' }\over{4f}}]({\p}B_2)^2+\nn\\
            &&2{\d}{\d}_1{\p}B_0.{\p}B_1
            +2{\d}{\d}_2{\p}B_0.{\p}B_2+
            2({\d}_1{\d}_2-{{qf' }\over{4f}}){\p}B_1.{\p}B_2+  \nn\\
            &&{{f' }\over f}{\p}q.[{\d}{\p}B_0+({\d}_1-
            1/2){\p}B_1+({\d}_2-1/2){\p}B_2]+{{f'' }\over{4f}}({\p}q)^2
\label{A.20} \eea
Using these in (\ref{A.12}) and including the dilaton term, 
a final expression for ${\cL}_G$ as in (\ref{4.4}) results.\\

\renewcommand{\theequation}{B.\arabic{equation}}
\setcounter{equation}{0}
{\large{\bf B. Analysis of the Diophantine equation for intersections}}\\
As we have seen in section 1,  possible marginal 
intersections of super p-branes
are governed by a Diophantine equation written as (\ref{1.27}) or
\be       (2{\Db}-d_1{\dt}_1)(2{\Db}-d_2{\dt}_2)=(d_1d_2-{\d}{\Db})^2
\label{B.1} \ee
There are at least two means for classifying the solutions of this equation.
Given the spacetime dimension $D$,  we can specify:\\ 
1) the number of common directions 
$({\d}-1)$  or  2) the number of angles $m$\\
and look for the possible dimensions of the branes $(d_1-1, d_2-1)$. 
We first present the method of analysis for arbitrary $D$'s below, 
and at the end summarize the results for interesting dimensions
$D=4, 6, 10, 11$. But before,  two simple cases may be distinguished:\\
{\bf$a$) The same-type branes}\\
This corresponds to: $d_1=d_2$ \& ${\a}_1={\a}_2$
so that ${\a}_1{\a}_2\geq 0$ and (\ref{1.21}) implies:
$2{\Db}-d_1{\dt}_1=+(d_1^2-{\d}{\Db})$ , i.e.
\be       {\d}=d_1-2
\label{B.2} \ee
This means that for two branes of the  `same type'  to marginally bind,  
all their angles except two of them must be vanishing (i.e. $m=2$). A result 
which is
in section 4 of this paper.\\
{\bf $b$) The self-dual pair of branes}\\
This case corresponds to: $d_1={\dt}_2$ \& ${\a}_1=-{\a}_2$
so that ${\a}_1{\a}_2\leq 0$ and (\ref{1.21}) implies:
$2{\Db}-d_1{\dt}_1=-(d_1{\dt}_1-{\d}{\Db})$ , i.e.
\be               {\d}=2
\label{B.3} \ee
This means that for a  `self-dual'  pair of branes to
marginally bind,  they must be intersecting over a string.\\
                                                                       
{\bf General restrictions}\\
Not all the solutions of (\ref{B.1}) are physically acceptable,  as we 
have two sets of restrictions:\\
First,  the definitions of dimensions and the condition of asymptotic 
flatness require                                    
\bea   &&0<{\d}\leq{\dl}<{\Db}\;\; ;\;\;{\l}=1, 2\nn\\
       &&0<d_1+d_2-{\d}<{\Db}
\label{B.4} \eea
Second,  the reality of $({\a}_1, {\a}_2)$ and the restriction on 
their relative sign require
\bea    &&{\dl}{\dtl}\leq2{\Db}\;\;;\;\;{\l}=1, 2\nn\\
         &&sgn(d_1d_2-{\d}{\Db})=sgn({\a}_1{\a}_2)
\label{B.5} \eea

{\bf The method of classification by ${\d}$}\\
Noting the symmetry of (\ref{B.1}) relative to $(d_1, d_2)$,  
we introduce the new 
variables
\be      r=d_1d_2\;\;\;\;\;\; ,  \;\;\;\;\;\;s=d_1+d_2
\label{B.6} \ee
in terms of which (\ref{B.1}) is written as
\be       r=2s+P+{R\over{s-Q}}
\label{B.7} \ee
where $(P, Q, R)$ are integers defined as
\be       P=4({\d}-2)\;\;\; , \;\;\;
Q={\Db}+2({\d}-2)\;\;\; , \;\;\;R=(8-{\Db})({\d}-2)^2
\label{B.8} \ee
Therefore solving (\ref{B.1}) 
for $(d_1, d_2)$ (assuming $({\d}, {\Db})$ are given) is 
equivalent to solving (\ref{B.7}) 
for $(r, s)$ generally,  and selecting then those
solutions for which $(d_1, d_2)$ are integers,  i.e.
\bea
         s^2-4r=\left\{\begin{array}{lll}
                 (2q)^2\;\; &, &\;\;s\in 2\bf Z \\
                 (2q+1)^2\;\; &, &\;\;s\in 2\bf Z+1
                 \end{array}
                \;\;\;\;\; (q\in\bf Z)  \right.
\label{B.9} \eea
On the other hand (\ref{B.7}) by 
itself has a finite set of solutions for $(r, s)$, 
for given values of $(P, Q, R)$,  which 
may be obtained by demanding that 
the integer $(s-Q)$ enumerates the integer $R\neq 0$.\\

{\it The $R=0$ cases:}\\
These include: $D=10$  and/or  ${\d}=2$. 
In both cases each of the relations: 
$s=Q$  ,   $r=2s+P$ solves the equation (\ref{B.7}). 
The corresponding solutions
for $(d_1, d_2)$ then are given by
\bea
          D=10\;\; :\;\;\left\{ \begin{array}{lll}
                          &&d_1+d_2=2({\d}+2)\\
                           or\\
                          &&(d_1-2)(d_2-2)=4({\d}-1)
                          \end{array}
                          \right.
\label{B.10} \eea

\bea
          {\d}=2\;\; :\;\;\left\{ \begin{array}{lll}
                          &&d_1+d_2={\Db}\\
                           or\\
                          &&(d_1, d_2)=(4, 4), (3, 6)
                          \end{array}
                          \right.
\label{B.11} \eea

{\bf The method of classification by $m$}\\
For simplicity we may assume: $d_1\leq d_2$ and thus ${\d}=d_1-m$. So noting
(\ref{B.6}) and (\ref{B.8}),  we can write (\ref{B.7}) as
\be    d_1d_2=2(d_1+d_2)+4(d_1-m-2)
+{{(8-{\Db})(d_1-m-2)^2}\over{(d_2-d_1)-{\Db}+2(m+2)}}
\label{B.12} \ee
Defining the 1 to 1 map: $(d_1, d_2)\mapsto (x, y)$ 
in the domain of $\bf Z$ by
\be          x=d_1-(m+2)\;\;\; , \;\;\; y=(d_2-d_1)-{\Db}+2(m+2)
\label{B.13} \ee
equation (\ref{B.12}) in $(x, y)$ variables transforms to
\be          x^2+yx+(my-k-{l\over{y-n}})=0
\label{B.14} \ee
where $(n, k, l)$ are integers defined by
\be   n{\;\=\;}8-{\Db}\;\; , \;\; k{\;\=\;} (m-2)^2\;\; , \;\;l{\;\=\;}nk
\label{B.15} \ee
Solving (\ref{B.1}) for $(d_1, d_2)$ (assuming $(D, d_1-{\d})$ as given) 
is equivalent
to solving (\ref{B.14}) for $(x, y)$ which is much easier. 
In fact equation (\ref{B.14})
(as (\ref{B.7})) has a finite set of solutions for given values of
$(m, n, k, l)$,  which can be found easily by 
demanding that the integer $(y-n)$
must enumerate the integer $l\neq 0$ 
(for $l=0$ see below),  and further that
\bea      
          y^2-4(my-k-{l\over{y-n}})=\left\{ \begin{array}{lll}
                 (2q)^2\;\; &, &\;\;s\in 2\bf Z \\
                 (2q+1)^2\;\; &, &\;\;s\in 2\bf Z+1
                 \end{array}
                 \;\;\;\;(q\in\bf Z)  \right.
\label{B.16} \eea
so that (\ref{B.14}) has integer solutions for $x$.\\

{\it The $l=0$ cases:}\\
By (\ref{B.15}) these include: 
$D=10$  and/or  $m=2$. Again (like the $R=0$ case)
the solution in both cases lie on 
two branches, i.e.: $y=n$  ,   $y=-x+m-{{4(m-1)}\over{x+m}}$.
The corresponding solutions for $(d_1, d_2)$ then are given by
\bea
          D=10\;\; :\;\;\left\{ \begin{array}{lll}
                          &&d_2-d_1=4-2m\\
                          or\\
                          &&(d_1-2)(d_2-6)=4-4m
                          \end{array}
                          \right.
\label{B.17} \eea

\bea
          m=2\;\; :\;\;\left\{ \begin{array}{lll}
                          &&d_1=d_2\\
                          or\\
                          &&(d_1, d_2)=(3, D-8), (4, D-6), (6, D-5)
                          \end{array}
                          \right.
\label{B.18} \eea
where the condition $d_1\leq d_2$ in 
the second class of (\ref{B.18}) requires that $D\geq 11, 10, 11$
respectively. Note that the two classes: 
(\ref{B.17}) \& (\ref{B.10}) are the same. The condition
$d_1\leq d_2$ restricts the first 
class of the (\ref{B.17}) solutions to those with
$m=0, 1, 2$. In particular for two 
parallel branes ($m=0$) or two non-parallel branes
with only one angle ($m=1$) in this class,  
we obtain: $d_2-d_1=$4 or 2  respectively, 
in agreement with the results of \cite{36} 
for $D$-branes. In fact noting that the
dilaton coupling for a D- $(d-1)$-brane 
is ${\a}(d)={\pm} (4-d)/4$, with $+(-)$ sign
for branes with electric (magnetic) RR charges,  
we conclude that the first
(second) class of solutions 
in (\ref{B.17}) in the case of two
$D$-branes,  describes bound states of 
two $D$-branes of the same (opposite)
 `electromagnetic'  type.\\

{\bf Special cases of $D=4, 6, 10, 11$}\\
We present here the summary of the above classifications of solutions for
$D=4, 6, 10, 11$. In this summary we relax the 
restrictions: ${\d}>0$ ,  $d<{\Db}$
to the extent that: ${\d}\geq 0$ ,  $d<D$; so 
that the solutions include  `instanton-like' 
objects as well as the  `non-asymptotic flat'  configurations.
The last column in each table idicates that the two branes are of the same 
or different electric/magnetic type according to ${\a}_1{\a}_2$ to be
positve or negative.
$$
\begin{array}{|c|c|c|c|}\hline
\d&(d_1, d_2)&m&sgn({\a}_1{\a}_2)\\ \hline
0&(2, 2)&2&+ \\ \hline
\end{array} 
$$               
\begin{center}  table (2)  $D=4$  \end{center}
$$
\begin{array}{|c|c|c|c|}\hline
\d&(d_1, d_2)&m&sgn({\a}_1{\a}_2)\\ \hline
0&(2, 2)&2&+\\ 
1&(3, 3)&2&+ \\
2&(2, 2)&0&- \\ \hline
\end{array} 
$$               
\begin{center}  table (3)  $D=6$  \end{center}
$$
\begin{array}{|c|c|c|c|}\hline
\d&(d_1, d_2)&m&sgn({\a}_1{\a}_2)\\ \hline
0&(0, 4), (1, 3), (2, 2), (1, 6)&0, 1, 2, 1&0,+,+,+ \\ 
1&(1, 5), (3, 3), (2, 2), (2, 3), ..., (2, 9)&0, 2, 1, 1, ..., 1
&-,+,-,-,0,+,...,+ \\
2&(2, 6), (3, 5), (4, 4), (3, 6)&0, 1, 2, 1&-,-,0,+\\
3&(3, 7), (4, 6), (5, 5)&0, 1, 2&-,0,+ \\
4&(4, 8), (5, 7), (6, 6), (5, 6)&0, 1, 2, 1&0,+,+,- \\
5&(5, 9), (6, 8), (7, 7), (6, 6)&0, 1, 2, 1&+,+,+,- \\
6&(6, 7)&0&- \\ \hline
\end{array} 
$$               
\begin{center}  table (4)  $D=10$  \end{center}
$$
\begin{array}{|c|c|c|c|}\hline
\d&(d_1, d_2)&m&sgn({\a}_1{\a}_2)\\ \hline
0&(0, 3), (2, 2), (0, 6)&0, 2, 0&0,+,0 \\ 
1&(3, 3)&2&0 \\
2&(2, 7), (3, 6)&0, 1&-,0 \\
3&(3, 9)&0&0 \\
4&(6, 6)&2&0 \\
5&(7, 7)&2& +\\
6&(6, 9), (8, 8)&0, 2&0,+ \\ \hline
\end{array} 
$$               
\begin{center}  table (5)  $D=11$  \end{center}

\renewcommand{\theequation}{C.\arabic{equation}}
\setcounter{equation}{0}
{\large{\bf C. Definition of the angles between two branes }}\\
We have seen at the beginning of 
section 4 that the  `relative'  orientation of a
pair of branes can be described in 
terms of the parameters ${\gmmp}$ entering
the metric as in (\ref{4.1}). The asymptotic form of this metric, 
restricted to the subspace of 
coordinates $(y_1^m, y_2^{m' })$ then is written
as \ft{Refer to table (1),  section 1,  for 
the definitions of coordinates and subspaces.}
\be     d{\s}^2{\;\=\;}({\dy})^2+({\dyp})^2+2{\gmmp}{\dy}{\dyp}
\label{C.1} \ee
Therefore ${\gmmp}$'s are related to the angles between coordinates as
\be       {\gmmp}={cos ({\p}_m, {\p}_{m' })|}_{z{\ra}\infty}
\label{C.2} \ee
Clearly the $({\d}_1{\d}_2)$ numbers ${\gmmp}$ depend on the choice of the
(Cartesian) coordinate system: $(y_1^m, y_2^{m' })$,  
transforming as components of a rank (1, 1)
tensor under the rotations: 
$SO({\d}_1)\times SO({\d}_2)$ of these coordinates.
Therefore we can $not$ identify $({\gmmp})$ as the set of
$independent$ parameters needed to describe 
the relative orientation of the
two branes,  since they are related through 
these rotations. To do this,  we need
to specify the $maximal$ set of 
$SO({\d}_1)\times SO({\d}_2)-$invariant quantities.
We call such invariant parameters as 
the  `geometric'  or  `intrinsic'  angles of the
two branes. To give a simple 
description of these angles,  we use our intuitions
in 3-dimensional Euclidean 
geometry\ft {You may consider e.g. a line and a plane
at angle in the 3-space.}. 
We first take a pair of $arbitrary$ $unit$ vectors
$(\bf n_1, \bf n_2)$ within the 
$({\d}_1, {\d}_2)$ subspaces respectively as
\bea      
{\bf n}_1&=&\o_1^m{\p}_m\;\;\;\;\;\;\; ;\;\;\;\;\;\;
(\o_1^m)^2=1\nn\\
{\bf n}_2&=&\o_2^{m' }{\p}_{m' }\;\;\;\;\;\; ;\;\;\;\;\;\;
(\o_2^{m' })^2=1
\label{C.3} \eea
The angle between $(\bf n_1, \bf n_2)$ 
according to the metric (\ref{C.1}) then is
defined by
\be       {\ct}={\bf n}_1.{\bf n}_2={\gmmp}\o_1^m\o_2^{m' }
\label{C.4} \ee
We now define the  `geometric'  angles between the two branes to be
the  `non-trivial'  extremums of the 
quantity ${\l}{\;\=\;}{\ct}$. By  `non-trivial'  here,  we
mean those extremums which are not zero 
identically,  and not related together
by a change of signs. These extremums are 
obtained by extremizing the function
\be   S(\o )=\gmmp\o_1^m\o_2^{m' }
-{{{\l}_1}\over 2}(\o_1^m)^2-{{{\l}_2}\over 2}(\o_2^{m' })^2
\label{C.5} \ee
subject to the constraints in (\ref{C.3}) 
with $({\l}_1, {\l}_2)$ as the Lagrange
multipliers. For the extremum points we have
\be     {\l}_1={\l}_2={\l}=cos{\th}
\label{C.6} \ee
where ${\l}$ is obtained from the eigen-value secular equation
\be      det[\hat{\bga}-(1+{\l})\bf 1]=0
\label{C.7} \ee
where $\hat{\bga}\;{\;\=\;}\;\pmatrix{1&{\bga}
\cr {\bga}^T&1}$ is a $({\d}_1+{\d}_2)\times ({\d}_1+{\d}_2)$
matrix representing the metric 
tensor of (\ref{C.1}). So (\ref{C.7}) is written as
a polynomial equation in ${\l}$ of 
degree $({\d}_1+{\d}_2)$,  whose  `non-trivial'  roots
$\{{\l}_r\}_{r=1, ..., m}$ give, 
via ${\l}_r=cos{\th}_r$, the geometric angles $\{{\th}_r\}$. 
Not that by Hermiticity of $\bga$ 
all the ${\l}_r$'s are real. Also by
positive-definite ness of $d{\s}^2$ 
we can show that always $-1\leq {\l}_r\leq 1$.
Further the secular equation (\ref{C.7}) can be written as 
\be           {\l}^{{\d}_2-{\d}_1}det ({\bga}{\bga}^T-{\l}^2)=0
\label{C.8} \ee
which (assuming ${\d}_1\leq {\d}_2$) 
implies that the roots of (\ref{C.7}) consist of
a set of $({\d}_2-{\d}_1)$ zeros and 
${\d}_1$ pairs of opposite numbers (this had been
expected, since reversing the 
direction of ${\bf n}_1$ or ${\bf n}_2$ in (\ref{C.4})
changes the sign of $cos {\th}$ but 
preserves its extremum property). So the number
of ${\th}_r$'s equals the degree of the 
determinant in (\ref{C.8}) as
a polynomial function of ${\l}^2$,  which is in general
\be       m=Min\{{\d}_1, {\d}_2\}
\label{C.9} \ee
This number is in fact the (maximum) number of the successive
rotations,  required
for bringing the smaller in dimension 
brane from a parallel to an angled status
relative to the other brane.\\
   
\renewcommand{\theequation}{D.\arabic{equation}}
\setcounter{equation}{0}
{\large{\bf D. Derivation of useful $H$-surface identities}}\\
The model for the form-field sector of 
the RL (\ref{3.47}) can be simulated by a 
simplified model in the discrete mechanics as
\be
L[x, y]=-\2 g_{ij} (x){\dot y}^i{\dot y}^j
\label{D.1} \ee
where $(x^\a )$ and $(y^i)$ are two sets of 
dynamical variables and $g_{ij} (x)$
is any invertible  `metric tensor' . 
The \eoms for $y^i$'s are obviously integrable and
yield
\be
{\dot y}^i =c_jg^{ij} (x)
\label{D.2} \ee
where $c_j$'s are some integration constants. 
Eliminating $y^i$'s in (\ref{D.1})
by this equation,  we obtain
\be
L[x]=-\2 c_ic_jg^{ij} (x)
\label{D.3} \ee
Applying equation (\ref{D.2}) again,  it 
is easy to prove the  `on-shell identity' :
\be
{\p\over{\p x^\a}}L[x, y]=-{\p\over{\p x^\a}}L[x]
\label{D.4} \ee
Changing the role of $(x^\a , y^i, \g_{ij} , L)$ to
$\left (\fa , \Ari , e^{2{\a}_r\vf}\hp , \LF \right )$ 
in this equation,  and using
the definitions (\ref{Ul}) and (\ref{3.57}),  
we obtain the  `H-surface identity' :
\be
U_{{\l}{\lp} , {\a}}({\bff} , {\bH} )=-u_{{\l}{\lp} , {\a}}({\bff} )
\label{D.5} \ee
Differentiating the $H$-surface 
identity (\ref{3.56}) relative to $H_{\ls}$ and using
(\ref{D.5}),  we obtain another identity:
\be
U_{\l\lp , \ls}(\bff , H)=2\pls\ul (\bff )=2\pls\Ul (\bff ,H )
\label{D.6} \ee
Using this result in the equation (\ref{3.54}),  
after simple manipulations we obtain
\be
\plp U_{\l\ls}=\pls U_{\l\lp}
\label{D.7} \ee
Noting the symmetry of $\Ul$ in its two 
indices,  this equation implies the
existence of a function $U(H)$,  such that
\be
\Ul (\bff (H), H)=\pl\plp U(H)
\label{D.8} \ee
This result could be seen in a (somehow) different manner. Let's define
\be
U^r_{\l\lp}\;\=\; e^{-2{\a}_r\vf}\hp \pl\Ari (H)\plp\Arj (H)
\label{D.9} \ee
so that $\Ul =\sum_{r=1}^nU^r_{\l\lp}$ by (\ref{Ul}). 
From (\ref{3.55}) it is evident that
\be
U^r_{\l\lp}(\bff (H), H)=\pl (c^{r(i)}_{\lp}\Ari )=\plp (c^{r(i)}_{\l}\Ari )
\label{D.10} \ee
This implies the existence a function $U^r(H)$,  
in terms of which we can write
\bea
&&\clri \Ari =\pl U^r\\
&&U^r_{\l\lp}=\pl\plp U^r
\label{D.11} \eea
which proves (\ref{D.8}) again.\\

\end{document}